\documentclass[12pt]{article}

\ifx\pdfoutput\undefined
\usepackage[dvips,bookmarks]{hyperref}
\else
\usepackage{hyperref}
\fi
\hypersetup{colorlinks=false,bookmarksopen,bookmarksnumbered,citecolor=blue,
   pdfstartview=FitH}

\usepackage{latexsym}

\usepackage{amssymb,amsfonts,amsmath}
\usepackage{graphicx} 
\usepackage{indentfirst}

 \usepackage{bbm}

\parskip=\medskipamount

\newcommand {\cC}{{\cal C}}
\newcommand {\cD}{{\cal D}}

\newcommand {\cF}{{\cal F}}

\newcommand {\cJ}{{\cal J}}
\newcommand {\cK}{{\cal K}}
\newcommand {\cL}{{\cal L}}
\newcommand {\cM}{{\cal M}}
\newcommand {\cN}{{\cal N}}
\newcommand {\cO}{{\cal O}}

\newcommand {\cR}{{\cal R}}

\newcommand {\cT}{{\cal T}}

\newcommand {\cV}{{\cal V}}

\newcommand {\cX}{{\cal X}}

\newcommand{\bD}{{\bf D}}

\def\a{\alpha}

\def\b{\beta}
\def\c{\chi}
\def\d{\delta}
\def\e{\epsilon}

\def\g{\gamma}
\def\G{\Gamma}

\def\k{\kappa}
\def\l{\lambda}
\def\m{\mu}
\def\n{\nu}
\def\o{\omega}

\def\q{\theta}
\def\r{\rho}
\def\s{\sigma}

\def\x{\xi}
\def\z{\zeta}
\def\D{\Delta}
\def\F{\Phi}

\def\L{\Lambda}
\def\O{\Omega}

\def\S{\Sigma}
\def\U{\Upsilon}
\def\X{\Xi}

\def\rd{{\rm d}}
\def\ri{{\rm i}}
\def\re{{\rm e}}

\newcommand{\ve}{\varepsilon}                            %new
\newcommand{\pa}{\partial}                           %new
\newcommand{\hf}{\frac12}
\newcommand{\vf}{\varphi}
\newcommand{\be}{\begin{equation}}
\newcommand{\ee}{\end{equation}}
\newcommand{\bea}{\begin{eqnarray}}
\newcommand{\eea}{\end{eqnarray}}
\newcommand{\non}{\nonumber}
\newcommand{\1}{\underline{1}}
\newcommand{\2}{\underline{2}}

\newcommand{\bm}[1]{\mbox{\boldmath$#1$}}

\def\double #1{#1{\hbox{\kern-2pt $#1$}}}

\newcommand{\hm}{{\hat{m}}}

\newcommand{\ha}{{\hat{a}}}
\newcommand{\hb}{{\hat{b}}}
\newcommand{\hc}{{\hat{c}}}
\newcommand{\hd}{{\hat{d}}}
\newcommand{\he}{{\hat{e}}}

\newcommand{\hA}{{\hat{A}}}

\newcommand{\hal}{{\hat{\a}}}
\newcommand{\hbe}{{\hat{\b}}}
\newcommand{\hga}{{\hat{\g}}}
\newcommand{\hde}{{\hat{\d}}}

\newcommand{\bsubeq}{\begin{subequations}}
\newcommand{\esubeq}{\end{subequations}}

\begin{document}
%%%%%%%%%%%%%%%%
%%%%%%%%%%%%%%%%
\begin{titlepage}
\begin{flushright}
June, 2014\\
\end{flushright}

\begin{center}
{\Large \bf 
Symmetries of curved superspace in five dimensions}
\end{center}

\begin{center}

{\bf
Sergei M. Kuzenko,
Joseph Novak and Gabriele Tartaglino-Mazzucchelli
} \\
\vspace{5mm}

\footnotesize{
{\it School of Physics M013, The University of Western Australia\\
35 Stirling Highway, Crawley W.A. 6009, Australia}}  
~\\
\texttt{joseph.novak,\,gabriele.tartaglino-mazzucchelli@uwa.edu.au}\\
\vspace{2mm}

\end{center}

\begin{abstract}
\baselineskip=14pt
We develop a formalism to construct supersymmetric backgrounds
within the superspace formulation for five-dimensional (5D) conformal supergravity 
given in arXiv:0802.3953. Our approach is applicable to any off-shell formulation 
for 5D minimal Poincar\'e and anti-de Sitter  supergravity theories realized 
as the Weyl multiplet coupled with two compensators. 
For those superspace backgrounds which obey the equations of motion for 
(gauged) supergravity, we naturally reproduce the supersymmetric solutions constructed
a decade ago by Gauntlett et al.  For certain supersymmetric backgrounds 
with eight supercharges, we construct a large family of off-shell supersymmetric sigma models 
such that the superfield Lagrangian is given in terms of the K\"ahler potential of a real analytic 
K\"ahler manifold. 
\end{abstract}

\vfill
\end{titlepage}

\newpage
\renewcommand{\thefootnote}{\arabic{footnote}}
\setcounter{footnote}{0}

\tableofcontents{}
\vspace{1cm}
\bigskip\hrule

%%%%%%%%%%%%%%%%%%%%%%%%%%%%%%%%%%%%%%%%%%%%%%%%%%%%%%

\section{Introduction}
\setcounter{equation}{0}

Six years ago, two of us 
developed the 
superspace approach to off-shell $\cN=1$ supergravity-matter couplings 
in five dimensions (5D)  \cite{KT-Msugra5D,KT-Msugra5D2,KT-Mconfsugra5D}.\footnote{In
 five dimensions, different authors use different notations, 
$\cN=1$ or $\cN=2$, for supersymmetric theories with eight supercharges.  
The notation $\cN=1$ is used, e.g., in Refs.  \cite{KT-Msugra5D,KT-Msugra5D2,KT-Mconfsugra5D}.
The rationale for its use
is that the case of eight supercharges corresponds to simple supersymmetry. 
The alternative notation $\cN=2$ is used, e.g., in \cite{Zucker,Ohashi,Bergshoeff}.
 The reason for 
this choice is that dimensional reduction of five-dimensional theories with eight 
supercharges leads to $\cN=2$ theories in four dimensions.} 
As concerns the Weyl multiplet of 5D conformal supergravity, 
its formulation given in \cite{KT-Mconfsugra5D} may be simply 
thought of  as  an alternative realization of the one discovered a few years earlier within the 
component superconformal tensor calculus \cite{Ohashi,Bergshoeff}.\footnote{The
minimal  multiplet of 5D $\cN=1$ supergravity was originally sketched,  
within a superspace setting, by Howe in 1981 \cite{Howe5Dsugra}
(using the supercurrent multiplet constructed in \cite{HL}) and fully elaborated in 
 \cite{KT-Msugra5D,KT-Msugra5D2}.
It was re-discoverd by Zucker \cite{Zucker} 
who elaborated on the component implications of \cite{Howe5Dsugra}.}
However, the real power of the superspace approach of
\cite{KT-Msugra5D,KT-Msugra5D2,KT-Mconfsugra5D}
is that it offers a generating formalism 
to realize the most general locally supersymmetric $\sigma$-model couplings 
and hence, in principle, to construct new quaternionic K\"ahler metrics.
This is achieved by making use of the concept of covariant projective supermultiplets
\cite{KT-Msugra5D,KT-Msugra5D2,KT-Mconfsugra5D}.
These supermultiplets are a curved-superspace extension of  the so-called superconformal 
projective multiplets \cite{K2006}, which in the 4D $\cN=2$ super-Poincar\'e case reduce 
to the off-shell projective multiplets pioneered by Lindstr\"om and Ro\v{c}ek \cite{LR}.
Among the most interesting covariant projective supermultiplets are polar ones 
that have infinitely many auxiliary fields. Such off-shell supermultiplets are practically impossible 
to  engineer or to deal with in the framework of  superconformal tensor calculus. 
This is why they had never appeared 
within the component settings of  \cite{Ohashi,Bergshoeff,Zucker}, which deal only with 
hypermultiplets either with a gauged central charge \cite{Ohashi,Zucker} 
or that are on-shell \cite{Bergshoeff}.

The superspace formulation developed in  \cite{KT-Msugra5D,KT-Msugra5D2,KT-Mconfsugra5D}
provides a universal setting
to generate off-shell supersymmetric field theories on 
 curved backgrounds. For instance, the general 5D $\cN=1$ rigid supersymmetric theories 
in AdS$_5$, which were constructed in \cite{KT-M07}, can easily be read off from 
the supergravity-matter systems proposed in \cite{KT-Msugra5D,KT-Msugra5D2,KT-Mconfsugra5D}
by properly freezing the supergravity fields. 
Of course, the problem of constructing  supersymmetric field theories on a given spacetime
is well formulated only if this manifold is a supersymmetric background, i.e. it admits rigid supersymmetries. 
Thus one is naturally led to the more general problem of looking for those curved superspaces that
possess (conformal) isometries. In the case of 4D $\cN=1$ old minimal supergravity, 
the latter problem was addressed in \cite{BK} almost twenty years ago. 
 The approach presented in \cite{BK} is  universal, for  
in principle it may  be generalized to supersymmetric backgrounds associated with any supergravity theory 
formulated in superspace.
In particular, it has already been used to construct rigid supersymmetric field theories
in 5D $\cN=1$ \cite{KT-M07}, 4D $\cN=2$ \cite{KT-M08,BKads,BKLT-M}
and 3D $(p,q)$ anti-de Sitter  \cite{KT-M11,KLT-M12,BKT-M}  superspaces. 

Recently, a number of publications have appeared devoted to the construction 
of supersymmetric backgrounds associated with  off-shell supergravity theories 
in diverse dimensions, see \cite{FS,Jia:2011hw,Samtleben:2012gy,Klare:2012gn,DFS,CDFKS1,CDFKS2,KMTZ,Liu:2012bi,Dumitrescu:2012at,Kehagias:2012fh,Festuccia3D,HTZ,deMH1,deMH2,Kom13,DKSS2,ClossetCremonesi} and references therein.   
Inspired by \cite{FS}, these works used component field considerations. 
In the case of 4D $\cN=1$ supergravity, it was shown 
\cite{K13} how to derive the key component results of, e.g.,  \cite{FS,KMTZ} from the 
more general superspace construction of \cite{BK}.
Recently, the formalism of \cite{K13} was extended to construct supersymmetric backgrounds \cite{KLRST-M} 
associated with all known off-shell formulations for 
3D $\cN=2$ supergravity  \cite{KT-M11,KLT-M11}. 
The results obtained are in agreement with the component considerations of 
\cite{Festuccia3D,HTZ,DKSS2}. In the present paper, we apply the ideas and techniques
developed in \cite{K13,KLRST-M}  to construct supersymmetric backgrounds  associated with 
5D $\cN=1$ supergravity. 

This paper is organized as follows. Section 2 contains a brief review of the superspace
formulation for 5D conformal supergravity \cite{KT-Mconfsugra5D}.
In section 3 we study (conformal) isometries of a background superspace. 
In section 4 we study bosonic backgrounds that possess at least one
(conformal) Killing spinor. Maximally supersymmetric backgrounds 
are described in section 5. Sections 6 and 7 are  concerned with additional restrictions 
on the background geometry, which arise when a single conformal compensator,
a vector multiplet or an $\cO(2)$ multiplet,  is turned on. 
Section 8 is devoted to supersymmetric backgrounds in off-shell supergravity. 
Supersymmetric solutions in Poincar\'e and anti-de Sitter supergravity theories
are studied in section 9.   Finally, concluding comments are given in section 10. 

The main body of the paper is accompanied by two technical appendices. 
In Appendix A we recall how the problem of computing the (conformal) isometries 
of a curved spacetime is addressed within the Weyl-invariant formulation for gravity. 
In Appendix B we discuss the properties of bilinears constructed from a conformal 
Killing spinor. 

%%%%%%%%%%%%%%%%%%%%%%%%%%%%%%%%%%%%%%%%%%%%%%%%%%
%%%%%%%%%%%%%%%%%%%%%%%%%%%%%%%%%%%%%%%%%%%%%%%%%%

\section{The Weyl multiplet in superspace}
\setcounter{equation}{0}

In this section we briefly review the superspace description  \cite{KT-Mconfsugra5D}
of the Weyl multiplet of 5D conformal supergravity. 
Our notation and conventions follow those introduced
in \cite{KL} (see also the appendix of \cite{KT-Msugra5D2}).

Let $z^{\hat{M}}=(x^{\hm},\q^{\hat{\mu}}_i)$
be local bosonic ($x$) and fermionic ($\q$) 
coordinates parametrizing  a curved five-dimensional  superspace
$\cM^{5|8}$,
where $\hm=0,1,\cdots,4$, $\hat{\mu}=1,\cdots,4$, and  $i=\1,\2$.
The Grassmann variables $\q^{\hat{\mu}}_i$
are assumed to obey the standard pseudo-Majorana reality condition
$(\q^{\hat{\mu}}_i)^* = \q_{\hat{\mu}}^i =\ve_{\hat{\m} \hat{\n}}\,  \ve^{ij} \, \q^{\hat{\nu}}_j  $.
The tangent-space group
is chosen to be  ${\rm SO}(4,1)\times {\rm SU}(2)$
and the superspace  covariant derivatives 
$\cD_{\hat{A}} =(\cD_{\hat{a}}, \cD_{\hat{\a}}^i)$
have the form 
\bea
\cD_{\hat{A}}&=&
E_{\hat{A}} + \O_{\hat{A}} + \F_{\hat{A}}
~.
\label{CovDev}
\eea
Here $E_{\hat{A}}= E_{\hat{A}}{}^{\hat{M}}(z) \,\pa_{\hat{M}}$ is the (inverse) supervielbein, 
with $\pa_{\hat{M}}= \pa/ \pa z^{\hat{M}}$,
\bea
\O_{\hat{A} }= \hf \,\O_{\hat{A}}{}^{\hb\hc}\,M_{\hb\hc}
= \O_{\hat{A}}{}^{\hbe\hga}\,M_{\hbe\hga}~,\qquad 
M_{\ha\hb}=-M_{\hb\ha}~, \quad M_{\hal\hbe}=M_{\hbe\hal}
\eea
is the Lorentz connection,  and
\bea
\F_{\hat{A}} = \F^{~\,kl}_{\hat{A}}\,J_{kl}~, \qquad
J_{kl}=J_{lk}
\eea
is the SU(2) connection. 
The Lorentz generators with vector indices ($M_{\ha\hb}$) and spinor indices
($M_{\hal\hbe}$) are related to each other by the rule:
$M_{\ha\hb}=(\S_{\ha\hb})^{\hal\hbe}M_{\hal\hbe}$. 
The generators of ${\rm SO}(4,1)\times {\rm SU}(2)$
act on the covariant derivatives as follows:\footnote{The operation of
(anti-)symmetrization of $n$ indices 
is defined to involve a factor $(n!)^{-1}$.}
\bea
{[}J^{kl},\cD_{\hal}^i{]}
= \ve^{i(k} \cD^{l)}_{\hat \a}~,~~~
{[}M_{\hal\hbe},\cD_{\hga}^k{]}
=\ve_{\hga(\hal}\cD^k_{\hbe)}~,~~~
{[}M_{\ha\hb},\cD_{\hc}{]}
=2\eta_{\hc[\ha}\cD_{\hb]}~,
\label{generators}
\eea
where $J^{kl} =\ve^{ki}\ve^{lj} J_{ij}$.

The supergravity gauge group is generated by local transformations
of the form 
\be
\d_\cK \cD_{\hat{A}} =[ \cK, \cD_{\hat{A}} ]~,
\qquad \cK = \x^{\hat{C}}(z) \cD_{\hat{C}} +\hf K^{\hat c \hat d}(z) M_{\hat c \hat d}
+K^{kl}(z)J_{kl}
~,
\label{tau}
\ee
with all the gauge parameters 
obeying natural reality conditions but are  otherwise  arbitrary. 
Given a tensor superfield $U(z)$ (with its indices suppressed),
its transformation law under the supergravity gauge group is
\bea
\d_\cK U = \cK  \, U~.
\eea

By construction, the covariant derivatives have (anti-)commutation relations 
of the general  form 
\bea
{[}\cD_{\hat{A}},\cD_{\hat{B}}\}&=&T_{\hat{A}\hat{B}}{}^{\hat{C}}\cD_{\hat{C}}
+\hf R_{\hat{A}\hat{B}}{}^{\hat{c}\hat{d}}M_{\hat{c}\hat{d}}
+R_{\hat{A}\hat{B}}{}^{kl}J_{kl}
~,
\label{algebra}
\eea
where $T_{\hat{A}\hat{B}}{}^{\hat{C}}$ is the torsion, 
and $R_{\hat{A}\hat{B}}{}^{\hat{c}\hat{d}}$   and $R_{\hat{A}\hat{B}}{}^{kl}$ are 
the SO(4,1) and SU(2) curvature tensors, respectively.

To describe conformal supergravity, 
the covariant derivatives have to obey certain constraints 
\cite{KT-Mconfsugra5D}. 
Upon solving the Bianchi identities for the constraints imposed,  it can be shown that the 
covariant derivatives are characterized by the (anti-)commutation relations:
\begin{subequations}
\bea
\big\{ \cD_{\hal}^i , \cD_{\hbe}^j \big\} &=&-2 \ri \,\ve^{ij}\cD_{\hal\hbe}
-\ri \,\ve_{\hal\hbe}\ve^{ij}X^{\hc\hd}M_{\hc\hd}
+{\ri\over 4} \ve^{ij}\ve^{\ha\hb\hc\hd\he}(\G_\ha)_{\hal\hbe}N_{\hb\hc}M_{\hd\he}
\non\\
&&
-{\ri\over 2}\ve^{\ha\hb\hc\hd\he}(\S_{\ha\hb})_{\hal\hbe}C_{\hc}{}^{ij}M_{\hd\he}
+4\ri \,S^{ij}M_{\hal\hbe}
+3\ri \, \ve_{\hal\hbe}\ve^{ij}S^{kl}J_{kl}
\non\\
&&
-\ri \, \ve^{ij}C_{\hal\hbe}{}^{kl}J_{kl}
-4\ri\Big(X_{\hal\hbe}+N_{\hal\hbe}\Big)J^{ij}
~,
\label{covDev2spinor-} \\
{[}\cD_\ha,\cD_{\hbe}^j{]}&=&
{1\over 2} \Big(
(\Gamma_{\hat{a}})_{\hbe}{}^{\hga}S^j{}_k
- X_{\ha\hb}(\Gamma^{\hat{b}})_{\hbe}{}^{\hga} \d^j_k
-{1\over 4}\,\ve_{\ha\hb\hc\hd\he}N^{\hd\he}(\Sigma^{\hb\hc})_{\hbe}{}^{\hga}
\d^j_k
+ (\S_\ha{}^{\hb})_{\hbe}{}^{\hga}C_\hb{}^j{}_k
\Big)
\cD_{\hga}^k
\non\\
&&
-{\ri\over 2}\Big((\G_\ha)_{\hbe}{}^\hga T^{\hc\hd}{}_\hga^{j}
+2(\G^{[\hc})_{\hbe}{}^\hga T_{\ha}{}^{\hd]}{}_\hga^{j}
\Big)M_{\hc\hd}
\non\\
&&
+\Big(3\X_{\ha}{}_\hbe^{(k}\ve^{l)j}
-{1\over 3}\cC_{\ha}{}_{\hbe}^{(k}\ve^{l)j}
-{5\over 4}(\G_\ha)_{\hbe}{}^\hga\cF_\hga^{(k}\ve^{l)j}
+{1\over 4}(\G_\ha)_{\hbe}{}^\hga\cN_\hga^{(k}\ve^{l)j}
\non\\
&&~~~
+{1\over 8}(\G_\ha)_{\hbe}{}^\hga\cC_{\hga}{}^{jkl}
-{11\over 24}(\G_\ha)_{\hbe}{}^\hga\cC_{\hga}^{(k}\ve^{l)j}
\Big)
J_{kl}
~.
\label{covDev2spinor-2}
\eea
\end{subequations}
The algebra of covariant derivatives is given in terms of 
dimension-1 tensor superfields, $S^{ij}$, $X_{\ha\hb}$, 
$N_{\ha\hb}$ and $C_\ha{}^{ij}$, and their covariant derivatives. They  
possess the symmetry properties:
\bea
S^{ij}=S^{ji}~,\qquad X_{\ha\hb}=-X_{\hb\ha}~,\qquad N_{\ha\hb}=-N_{\hb\ha}~,
\qquad C_\ha{}^{ij}=C_\ha{}^{ji}~.
\label{2.9}
\eea
Their reality properties are
\be
\overline{S^{ij} } =S_{ij}~, \qquad 
\overline{X_{\ha\hb}} =X_{\ha\hb}~, \qquad 
\overline{N_{\ha\hb}} =N_{\ha\hb}~, \qquad
\overline{C_\ha{}^{ij} }=C_{\ha ij }~.
\ee
The torsion superfields \eqref{2.9}
enjoy some additional differential constraints 
that follow from the Bianchi identities.
In terms of 
the irreducible components of $\cD_\hga^kX_{\ha\hb}$ and
$\cD_\hga^kC_{\ha}{}^{ij}$
defined by
\begin{subequations}
\bea
\cD_\hga^kX_{\ha\hb}&=&
W_{\ha\hb\hga}{}^k
+2(\G_{{[}\ha})_{\hga}{}^{\hde}\X_{\hb{]}\hde}{}^k
+(\S_{\ha\hb})_\hga{}^\hde \cF_{\hde}{}^k~, 
\non
\\
&&(\G^\ha)_\hal{}^\hbe\X_{\ha\hbe}{}^i=(\G^\ha)_\hal{}^\hbe W_{\ha\hb\hbe}{}^i=0
~,
\label{X-irreducible}\\
\cD_\hga^kC_{\ha}{}^{ij}&=&
\cC_\ha{}_\hga{}^{ijk}
-{2\over 3}\cC_{\ha}{}_{\hga}^{(i}\ve^{j)k}
-{1\over 2}(\G_\ha)_\hga{}^{\hde}\cC_\hde{}^{ijk}
+{1\over 3}(\G_\ha)_\hga{}^{\hde}\cC_\hde^{(i}\ve^{j)k}~,
\non
\\
&&
\cC_\ha{}_\hga{}^{ijk}=\cC_\ha{}_\hga{}^{(ijk)}~,~~
\cC_\hde{}^{ijk}=\cC_\hde{}^{(ijk)}~,~~~
(\G^\ha)_\hal{}^\hbe\cC_\ha{}_\hbe{}^{ijk}=0~,
\eea
\end{subequations}
the dimension-3/2 Bianchi identities are: 
\begin{subequations}
\bea
\cD_\hga^kN_{\ha\hb}&=&
-W_{\ha\hb\hga}{}^k
+4(\G_{{[}\ha})_{\hga}{}^{\hde}\X_{\hb{]}\hde}{}^k
+(\S_{\ha\hb})_\hga{}^\hde \cN_{\hde}{}^k~,
\label{D-N} 
\\
\cC_\ha{}_\hga{}^{ijk}&=&0~,
\label{3/2D-C}
\\
\cD_\hga^kS^{ij}&=&
-{1\over 4}\cC_{\hga}{}^{ijk}
+{5\over 12}\cC_\hga^{(i}\ve^{j)k}
+{1\over 2}\Big(3\cF_\hga^{(i}+\cN_\hga^{(i}\Big)\ve^{j)k}
~.
\label{3/2DS}
\eea
\end{subequations}
The 
tensor $T_{\ha\hb}{}^\hga_k$ in \eqref{covDev2spinor-2}
is the dimension-3/2 torsion.
Its explicit form is
\bea
T_{\ha\hb}{}_\hga^k&=&
{\ri\over 2}\cD_\hga^k X_{\ha\hb}
-{\ri\over 6}(\G_{[\ha})_\hga{}^\hde\cC_{\hb]}{}_\hde^{k}
+{\ri\over 4}(\S_{\ha\hb})_\hga{}^\hde\cC_\hde^{k}~.
\label{dim-3/2-torsion}
\eea

The above superspace
geometry  describes conformal supergravity 
due to the fact that the algebra of covariant derivatives
 is invariant under infinitesimal super-Weyl transformations of the form
\begin{subequations} \label{sW}
\bea
\d_\s \cD_\hal^i&=&\hf\s\cD_\hal^i+2(\cD^{\hga i}\s)M_{\hga\hal}-3(\cD_{\hal k}\s)J^{ki}~,
\label{sW1} \\
\d_\s \cD_\ha&=&
\s\cD_\ha
+\frac{\ri}{2}(\G_\ha)^{\hga\hde}(\cD_{\hga}^{k}\s)\cD_{\hde k}
-(\cD^\hb\s)M_{\ha\hb}
+{\ri\over 8}(\G_\ha)^{\hga\hde}(\cD_\hga^{(k}\cD_{\hde}^{l)}\s)J_{kl}
\label{sW2}
~,
\eea
\end{subequations}
provided the
components of the torsion transform as follows:
\begin{subequations}
\bea
\d_\s S^{ij}&=&\s S^{ij}
+{\ri\over 4}\,\cD^{\hal (i}\cD_{\hal}^{ j)}\s~,
\label{s-Weyl-Sij}\\
\d_\s C_{\ha}{}^{ij}&=&\s C_{\ha}{}^{ij}
+\frac{\ri}{2}\,( \G_\ha)^{\hga\hde} \cD_{\hga}^{(i}\cD_{\hde}^{j)}\s~,
\label{C-var}\\
\d_\s X_{\ha\hb}&=&\s X_{\ha\hb}
-{\ri\over 4}\, (\S_{\ha\hb})^{\hal\hbe}\cD_\hal^k\cD_{\hbe k}\s~,
\label{s-Weyl-X}\\
\d_\s N_{\ha\hb}&=&\s N_{\ha\hb}
-\frac{\ri}{2} \,(\S_{\ha\hb})^{\hal\hbe} \cD_{\hal}^{k}\cD_{\hbe k}\s~,
\label{s-Wey-N}
\eea
\end{subequations}
with the parameter $\s(z)$ being an arbitrary  real scalar superfield.\footnote{The finite 
form for the super-Weyl transformations is given in \cite{KT-M5DCF}.
As  compared with \cite{KT-Mconfsugra5D,KT-M5DCF}, we have rescaled the super-Weyl parameter 
$\s \to \hf \s$.}   
It follows that the tensor
\be
W_{\ha \hb} := X_{\ha\hb} -\hf  N_{\ha\hb}
\ee
transforms homogeneously, 
\bea
\d_\s W_{\ha\hb}&=&\s W_{\ha\hb}~,
\eea
and hence is a superspace generalization of the Weyl tensor.

%%%%%%%%%%%%%%%%%%%%%%%%%%%%%%%%%%%%%%%%%%%%%%%%%%

In complete analogy with $\cN=2$ supergravity in four dimensions
(see, e.g., \cite{FVP} for a review), 5D $\cN=1$ Poincar\'e or anti-de Sitter supergravity 
theories are obtained by coupling the Weyl multiplet with two off-shell 
conformal compensators, one of which is (almost) invariably a vector multiplet. 
Conceptually, this approach is a natural extension of the Weyl-invariant 
formulation for gravity reviewed in Appendix A.

%%%%%%%%%%%%%%%%%%%%%%%%%%%%%%%%%%%%%%%%%%%%%%%
%%%%%%%%%%%%%%%%%%%%%%%%%%%%%%%%%%%%%%%%%%%%%%%
%%%%%%%%%%%%%%%%%%%%%%%%%%%%%%%%%%%%%%%%%%%%%%%

\section{(Conformal) isometries}
\setcounter{equation}{0}

Consider some background superspace $\cM^{5|8}$ such that its geometry 
is of the type described  in the previous section.
In order to formulate rigid superconformal or rigid supersymmetric field theories
on  $\cM^{5|8}$, one has to determine all (conformal) isometries of this superspace.  
This can be done similarly to the case of 4D $\cN=1$ supergravity
described in detail in \cite{BK} and elaborated in \cite{K13}.
A similar analysis in the case of 3D $\cN=2$ supergravity 
has recently been carried out in 
 \cite{KLRST-M}.

\subsection{Conformal isometries}

Let $\x = \x^{\hat{A}} E_{\hat{A}} = \x^\ha E_\ha + \x^\hal_i E_\hal^i  $ 
be a  real supervector field on $\cM^{5|8}$.
It  is called  conformal Killing 
if one can associate with $\x$ a supergravity gauge 
transformation \eqref{tau} and an infinitesimal  super-Weyl transformation
\eqref{sW}
such that their combined action 
\bea
\d:=\d_\cK + \d_\s
\eea
does not change the covariant derivatives,
\bea
\d \cD_{\hat{A}} =0~.
\label{3.1}
\eea
These conditions, which appeared for the first time in \cite{KT-Mconfsugra5D}, 
clearly imply that all the torsion and curvature tensors
are invariant under the transformation $\d$. 
One may see that it suffices to demand only 
the spinor condition $\d\cD_\hal^i =0 $ 
in order for \eqref{3.1} to hold. 
A short calculation gives
\bea
\d\cD_\hal^i&=&
\Big(
 \x^{\hat{C}}T_{\hat{C}}{}_{\hal}^i{}^\hbe_j
- \cD^i_\hal \x_j^{\hbe} 
 +K_\hal{}^{\hbe}\d^i_j
 +K^i{}_{j}\d_\hal^\hbe
+\hf\s\d_\hal^\hbe\d^i_j
\Big)\cD_\hbe^j
\non\\
&&
+\Big(
\x^{\hat{C}}T_{\hat{C}}{}_\hal^i{}^{\hb}
-\cD_\hal^i \x^{\hb}
\Big)\cD_{\hb}
\non\\
&&
+\Big(
\x^{\hat{D}}R_{\hat{D}}{}_\hal^i{}{}_{\hbe\hga}
 - \cD^i_\hal K_{\hbe\hga}
-2\ve_{\hal(\hbe}\cD_{\hga)}^{i}\s \Big)
 M^{\hbe\hga}
\non\\
&&
+\Big(
\x^{\hat{D}}R_{\hat{D}}{}_\hal^i{}^{jk}
-\cD^i_\hal K^{jk}
+3\ve^{i(j} \cD_{\hal}^{k)}\s
\Big)
J_{jk}
~.
\label{3.2}
\eea
The right-hand side of \eqref{3.2} is a  combination of the four 
 linearly independent operators $\cD_\hbe^j$, $\cD_\hb$, $M^{\hbe \hga}$ 
 and $J_{jk}$. 
 Requiring $\d\cD_\hal^i=0$ leads to four different equations.
Making use of the explicit form of the torsion,
the  equations associated with the operators $\cD_\hbe^j$ and $\cD_{\hb}$ 
in the right-hand side of \eqref{3.2} may be written as 
\bsubeq \label{3.3}
\bea
\cD^i_\hal \x^j_{\hbe}
&=&
\hf \x^{\ha}\Big(
(\Gamma_{\hat{a}})_{\hal\hbe}S^{ij}
+ X_{\ha\hb}(\Gamma^{\hat{b}})_{\hal\hbe} \ve^{ij}
+{1\over 4}\,\ve_{\ha\hb\hc\hd\he}N^{\hb\hc}(\Sigma^{\hd\he})_{\hal\hbe}\ve^{ij}
+ (\S_\ha{}^{\hb})_{\hal\hbe}C_\hb{}^{ij}
\Big)
\non\\
&&
 -K_{\hal\hbe}\ve^{ij}
 -K^{ij}\ve_{\hal\hbe}
+\hf\s\ve_{\hal\hbe}\ve^{ij}
~,
\label{3.3a}
\\
\cD_\hal^i \x_\hb
&=&
2\ri(\G_\hb)_{\hal}{}^{\hde}\x^i_{\hde}
~.
\label{3.3b}
\eea
\esubeq
After introducing $\x_{\hal\hbe}=(\G^\ha)_{\hal\hbe}\x_\ha$, 
equation \eqref{3.3b} is equivalent to
%%%%%%
\bea
\cD_\hal^i \x_{\hbe\hga}=
-8\ri\Big(
\x^i_{[\hbe}\ve_{\hat\g]\hat \a } 
+\frac{1}{4}\x^i_{\hal}\ve_{ \hbe \hga}
\Big)
~.
\eea
The relations \eqref{3.3} 
imply  that the  parameters $\x^\hal_i,\, K_{\hal\hbe},\,K^{ij}$ and $\s$ are uniquely expressed in terms 
of $\x^\ha$ and its covariant derivatives as follows:
\bsubeq
\label{3.4}
\bea
\x^i_{\hal}&=&
\frac{\ri}{10}(\G^\ha)_{\hal}{}^{\hbe}\cD_\hbe^{i} \x_\ha
~,
\label{3.4a}
\\
%%%
K_{\hal\hbe}&=&
\hf \cD^k_{(\hal} \x_{k\hbe)}
+ {1\over 8}\x^{\ha}\ve_{\ha\hb\hc\hd\he}N^{\hb\hc}(\Sigma^{\hd\he})_{\hal\hbe}
~,~~~~~~
\label{3.4b}
\\
%%%
K^{ij}&=&
\frac{1}{4} \cD^{(i}_{\hga} \x^{j)\hga}
=
\frac{\ri}{40}(\G^\hc)^{\hal\hbe}\cD^{(i}_{\hal}\cD_\hbe^{j)} \x_\hc
~,
\label{3.4c}
\\
%%%
\s&=&
\frac{1}{4}\cD^i_\hal \x_{i}^{\hal}
~.
\label{3.4d}
\eea
\esubeq
Since all the  parameters in $\cK$ 
and the super-Weyl parameter  $\s$ are functions of $\x$, we 
may use the notation $\cK = \cK [ \x ]$ and $\s = \s[ \x ]$.
It is important to note that equation \eqref{3.3b} implies a fundamental 
constraint on $\x^\ha$,
\bea
\Big(
\d_\hal{}^\hbe\d_\ha{}^\hb
+\frac{1}{5}(\G_\ha
\G^\hb)_{\hal}{}^{\hbe}
\Big)\cD_\hbe^{i} \x_\hb
=
\frac{4}{5}\Big(
\d_\hal{}^\hbe \d_\ha{}^\hb
-\hf(\S_\ha{}^\hb)_{\hal}{}^{\hbe}
\Big)\cD_\hbe^{i} \x_\hb = 0
~.
\label{3.5}
\eea
This equation means that the gamma-traceless component of 
the spin-vector $\cD^i_\hal \x_\hb$ is zero.
Eqs. \eqref{3.4} and  \eqref{3.5} imply the conformal Killing equation 
\bea
\cD_{(\ha}\x_{\hb)} 
&=&
\frac{1}{5}\eta_{\ha\hb}\cD_{\hc}\x^{\hc}~.
\eea
Other consequences of \eqref{3.4} and \eqref{3.5} are 
\bsubeq
\bea
\s &=&\frac{1}{5}\cD_{\hc}\x^{\hc}
~,
\\
 K_{\ha\hb}
&=&
\cD_{[\ha}\x_{\hb]} 
\quad \Longleftrightarrow \quad
K_{\hal\hbe}
=
\hf (\S^{\ha\hb})_{\hal\hbe}\cD_{\ha}\x_{\hb} 
~.
\eea
\esubeq

If eq.  \eqref{3.5} holds and the conditions \eqref{3.4} are adopted, 
it can be proved that equation \eqref{3.1} is identically satisfied.
 Therefore, \eqref{3.5} is the fundamental equation containing all the information 
about the conformal Killing supervector fields. 
This means that the conformal Killing supervector field
can alternatively be defined as a real supervector field,
\bea
\x = \x^{\hat{A}} E_{\hat{A}} ~, \qquad \x^{\hat{A}}  \equiv (\x^\ha , \x^\hal_i) =
 \Big( \x^\ha , \frac{\ri}{10}\cD_{ \hbe i} \x^{\hal\hbe}\Big) ~,
\eea
obeying the master equation \eqref{3.5}.

If $\x_1$ and $\x_2$ are two conformal Killing supervector fields, their 
Lie bracket $[\x_1, \x_2]$ is a conformal Killing supervector field. 
It is obvious that, for any real $c$-numbers $r_1$ and $r_2$, the linear combination
 $r_1 \x_1 + r_2 \x_2$ is a  conformal Killing supervector field. 
Thus the set of all conformal Killing supervector
fields is a super Lie algebra. The conformal Killing supervector fields of 
$\cM^{5|8}$
generate  symmetries 
of a superconformal field theory on this superspace. 

\allowdisplaybreaks{

We have not yet analysed the equations 
associated with the generators $M^{\hat \b \hat \g}$ and $J_{jk}$ in the right-hand side of \eqref{3.2}. They are
\begin{subequations} \label{3.11}
\bea
\cD^i_\hal K_{\hbe\hga} &=&
\x^{\hat{D}}R_{\hat{D}}{}_\hal^i{}_{\hbe\hga}
-2\ve_{\hal(\hbe}\cD_{\hga)}^{i}\s 
 ~, \\
\cD^i_\hal K^{jk} &=&
\x^{\hat{D}}R_{\hat{D}}{}_\hal^i{}^{jk}
+3\ve^{i(j} \cD_{\hal}^{k)}\s~.
\eea
\end{subequations}
The relations \eqref{3.3} tell us that any spinor covariant derivative of $\x^{\hat B}$ 
can be represented as a linear combination of the parameters 
$\U = ( \x^{\hat B}, K^{\hat \b \hat \g} , K^{jk} , \s)$. The relations \eqref{3.11}
also tell us that 
$\cD_\hal^i \U$
can be represented as a linear combination of $\U$ and $\cD_\hga^k \s$. 
It turns out that  $\cD_\ha \U$ may be represented as  
a linear combination of $\U$ and $\cD_{\hat C} \s$. 
To prove this claim, let us look at the conditions of invariance of 
the dimension-1 torsion superfields,
$\d S^{ij}=0$, $\d C_\ha^{ij}=0$, $\d X^{\ha\hb} =0$ and $\d N^{\ha\hb}=0$.
These conditions\footnote{The conditions \eqref{3.12} are not new constraints. 
They are  satisfied identically provided eq.  \eqref{3.1} holds. We should point out  that 
eqs.~\eqref{3.12c} and \eqref{3.12d} imply the invariance condition of the super-Weyl tensor, 
which is
$\x^{\hat{C}}\cD_{\hat{C}}W_{\ha\hb} -2 K_{[\ha}{}^{\hc}W_{\hb]\hc}
+\s W_{\ha\hb}=0$.}
are: 
\bsubeq \label{3.12}
\bea
\cD^{\hal (i}\cD_{\hal}^{ j)}\s
&=&
4\ri\x^{\hat{C}}\cD_{\hat{C}}S^{ij}
+8\ri K^{(i}{}_{k}S^{j)k}
+4\ri\s S^{ij}
~,\\
%%%
( \G_\ha)^{\hga\hde} \cD_{\hga}^{(i}\cD_{\hde}^{j)}\s
&=&
2\ri\x^{\hat{C}}\cD_{\hat{C}}C_{\ha}{}^{ij}
+2\ri K_\ha{}^{\hb} C_{\hb}{}^{ij}
+4\ri K^{(i}{}_kC_{\ha}{}^{j)k}
+\s C_{\ha}{}^{ij}
~,\\
%%%
(\S_{\ha\hb})^{\hal\hbe}\cD_\hal^k\cD_{\hbe k}\s
&=&
-4\ri\x^{\hat{C}}\cD_{\hat{C}}X_{\ha\hb}
+8\ri K_{[\ha}{}^{\hc}X_{\hb]\hc}
-4\ri\s X_{\ha\hb}
~, \label{3.12c}\\
%%%
(\S_{\ha\hb})^{\hal\hbe}\cD_\hal^k\cD_{\hbe k}\s
&=&
-2\ri\x^{\hat{C}}\cD_{\hat{C}}N_{\ha\hb}
+4\ri K_{[\ha}{}^{\hc}N_{\hb]\hc}
-2\ri\s N_{\ha\hb}
~. \label{3.12d}
\eea
\esubeq
These identities tell us that two spinor derivatives of $\s$ may be represented 
as a linear combination of $\U$ and $\cD_{\hat C} \s$. This confirms the above claim.
Furthermore, it is not hard to deduce from the above identities 
that $\cD_\hal^i \cD_{\hat B} \s$  may be represented 
as a linear combination of $\U$ and $\cD_{\hat C} \s$.
As a result, applying any number of covariant derivatives to $\U$  gives
 a linear combination of $\U$ and $\cD_{\hat C} \s$.
We conclude  that the super Lie algebra of the conformal Killing vector fields 
on $\cM^{5|8}$ is finite dimensional. 
The number of its even and odd generators cannot exceed those in 
the 5D superconformal algebra  ${\mathfrak f} (4)$.}

To study supersymmetry transformations at the component level, 
it is useful to spell out one of the implications of \eqref{3.1}
with $\hat{A}= \ha$. Specifically, we consider the equation $\d\cD_{\ha} =0$ 
and read off the part proportional to a linear combination of the 
spinor covariant derivatives $\cD_\hga^k$. 
The result is
\bea
0
&=&
\cD_\ha\x^{\hga}_k
-\hf \Big(S_k{}^l
 (\Gamma_{\hat{a}})^{\hga}{}_{\hde}
-X_{\ha\hb}(\Gamma^{\hat{b}})^\hga{}_{\hde} \d^l_k
+ \frac{1}{4}   \d^l_k \, \ve_{\ha\hb\hc\hd\he}N^{\hd\he}(\Sigma^{\hb\hc})^\hga{}_{\hde}
-(\S_\ha{}^{\hb})^\hga{}_{\hde}C_\hb{}_k{}^l
\Big)
\x^\hde_l
\non\\
&&
+\x^{\hat{b}}T_{\ha\hat{b}}{}^{\hga}_k
-\frac{\ri}{2}(\G_\ha)^{\hga\hde} \cD_{\hde k}\s
~.
\label{conf-Killing-0}
\eea
It should be mentioned that \eqref{conf-Killing-0} is not a new constraint. 
It is satisfied identically provided the spinor condition $\d\cD_\hal^i =0 $ holds.

%%%%%%%%%%%%%%%%%%%%%%%%%%
%%%%%%%%%%%%%%%%%%%%%%%%%

\subsection{Conformally related superspaces}

A superspace $\widetilde{\cM}^{5|8}$ is said to be 
conformally related to
$\cM^{5|8}$ if the covariant derivatives $\widetilde{\cD}_{\hat A} $
of  $\widetilde{\cM}^{5|8}$ are obtained from  $\cD_{\hat A} $ by a finite 
super-Weyl transformation \cite{KT-M5DCF}, 
\begin{subequations}\label{CRSG}
\bea
\widetilde{\cD}_\hal^i&=&
\re^{\hf \r}\Big(\cD_\hal^i+2(\cD^{\hbe i}\r)M_{\hal\hbe}-3(\cD_{\hal j}\r)J^{ij}\Big)~,
\label{D_hal-ConfFlat}
\\
\widetilde{\cD}_{\ha}&=&
\re^{\r}\Big{(}
\cD_\ha
+\frac{\ri}{2}(\G_\ha)^{\hga\hde}(\cD_\hga^k\r)\cD_{\hde k}
-(\cD^\hb\r)M_{\ha\hb}
+\frac{\ri}{ 8}(\G_\ha)^{\hga\hde}(\cD_{\hga}^{k}\cD_{\hde}^{l}\r)J_{kl}
\non\\
&&~~~~
+\frac{\ri}{8}\ve_{\ha\hb\hc\hd\he}(\S^{\hb\hc})_{\hga\hde}(\cD^{\hga k}\r)
(\cD^{\hde}_k\r)M^{\hd\he}
+  \frac{5\ri}{8}(\G_\ha)^{\hga\hde}(\cD_{\hga}^k\r)(\cD_{\hde}^l\r)J_{kl}
\Big{)}
~,~~~~~~
\label{D_a-ConfFlat-2}
\eea
\end{subequations}
for some super-Weyl parameter $\r$. 
The two superspaces $\widetilde{\cM}^{5|8}$ and $\cM^{5|8}$ 
prove to have the same conformal Killing supervector fields. 
Given such a vector field $\x = \x^{\hat{A}} E_{\hat{A}} = \widetilde{\x}^{\hat{A}} 
\widetilde{E}_{\hat{A}}$, it may be shown that 
\begin{subequations}
\bea
\cK [ \widetilde{\x} ] &:=&  \widetilde{\x}^{\hat{B}} 
\widetilde{\cD}_{\hat{B}} + \hf K^{\hb\hc}[\widetilde{\x}] M_{\hb\hc} 
+ K^{kl} [\widetilde{\x} ]J_{kl}= \cK[\x ]~, \\
\s[ \widetilde{\x} ] &=& \s[\x] - \x \r \ . 
\eea
\end{subequations}
This is similar to the 4D and 3D analyses in \cite{K13} and \cite{KLRST-M}, respectively.

%%%%%%%%%%%%%%%%%%%%%

\subsection{Isometries}
\label{Isometries}

In order to describe $\cN=1$ Poincar\'e or anti-de Sitter supergravity theories, 
the Weyl multiplet has to be coupled with two off-shell conformal compensators 
that will be symbolically denoted $\X$. 
In general, both compensators are Lorentz scalars and have non-zero super-Weyl weights
$w_\X \neq 0$,
\bea
\d_\s\X=w_\X\s\X~.
\label{GenCom}
\eea
They may transform
in nontrivial  representations of the SU(2) group, which we do not specify at the moment.
The compensators are required to be nowhere vanishing in the sense that 
the SU(2) scalars $|\X|^2 $ should be strictly positive. 
Different off-shell supergravity theories correspond to
different choices of $\X$.

The off-shell supergravity multiplet is completely described in terms of the following data:
(i) the  superspace geometry described in section 2; and (ii) the conformal compensators $\X$. 
Given a supergravity background, its isometries should preserve both of these inputs. 
This leads us to the concept of Killing supervector fields.

A conformal Killing supervector field $\x = \x^{\hat{A}} E_{\hat{A}}$  on $\cM^{5|8}$ 
 is said to be Killing if the following conditions hold:
\begin{subequations} \label{3.17}
\bea
\Big[ \x^{\hat{B}} \cD_{\hat{B}} + \hf K^{\hb\hc}[\x] M_{\hb\hc} 
+ K^{kl}[\x ]J_{kl}, \cD_\hA \Big]  + \d_{\s [\x]} \cD_\hA &=&0~ , \label{3.17a}
\\ 
\Big(\x^{\hat{B}} \cD_{\hat{B}}  + K^{kl}[\x]J_{kl}   + w_\X \s [\x] \Big) \X& =&0~. \label{3.15b}
\eea
\end{subequations}
Here the parameters $ K^{\hb\hc}[\x]$, $K^{kl} [\x]$ and $\s[\x]$ 
are defined as in \eqref{3.4}.
The set of all Killing supervector fields on $\cM^{5|8}$ is a super Lie algebra. 
The Killing supervector fields of $\cM^{5|8}$ generate the spacetime (super)symmetries 
of all rigid  supersymmetric  field theories on this superspace. 

The Killing equations \eqref{3.17} are super-Weyl invariant 
in the following sense.  Consider  a supergravity background 
$(\widetilde{\cD}_{\hat A}, \widetilde{\X})$ 
that is 
conformal  to 
$(\cD_{\hat A}, \X)$, where  $\widetilde{\cD}_{\hat A}$
is related to $\cD_{\hat A}$ according to \eqref{CRSG} and $\widetilde{\X}$ is 
\bea
\widetilde{\X} = \re^{w_\X\s }\X~.
\eea
Then the equations \eqref{3.17} have the same form once rewritten 
in terms of $(\widetilde{\cD}_{\hat A}, \widetilde{\X})$.

Using the compensators $\X$ we can always construct a Lorentz and SU(2) 
scalar superfield $\F = \F(\X)$,  which is an algebraic function of $\X$, 
 nowhere vanishing, and has 
a  nonzero super-Weyl weight $w_\F$, 
\bea
\d_\s \F = w_\F \s \F~.
\eea
We have shown that the Killing equations \eqref{3.17} are super-Weyl invariant.
Super-Weyl invariance may be used to  impose the gauge
\bea
\F=1~.
\label{Xgauge1}
\eea
Then the  equation 
\bea
\Big(\x^{\hat{B}} \cD_{\hat{B}}  + K^{kl}[\x]J_{kl}   + w_\F \s [\x] \Big) \F& =&0~,
\eea
which follows from the Killing equations \eqref{3.15b}, 
becomes
\bea
\s [\x]&=&0
~.
\label{constr-sigma}
\eea

The above consideration is analogous to that given in Appendix A for the (conformal) 
isometries of a curved spacetime. The only difference is that a single scalar compensator 
is used in the case of gravity, while two compensators are needed in order 
to realize Poincar\'e or anti-de Sitter supergravities.

\section{Supersymmetric backgrounds: General formalism} 
\setcounter{equation}{0}

Our analysis will be restricted to curved  backgrounds 
without covariant fermionic fields --  that is,
\bea
\cD_\hal^i S^{kl}|= 0~, \qquad 
\cD_\hal^i C_{\ha}{}^{kl}|= 0~, \qquad 
\cD_\hal^i X_{\ha\hb}|= 0~, \qquad 
\cD_\hal^i N_{\ha\hb}|= 0~. 
\label{Ddim1}
\eea
Here the bar-projection is defined as usual:  
\bea
U|:= U(x,\q)\big|_{\q =0}~,
\eea 
for any superfield $U(z)=U(x,\q)$. The bar-projection of the superspace covariant derivatives 
is defined similarly:
\bea
\cD_{\hat{A}}| = E_{\hat{A}}{}^{\hat{M}}| \,\pa_{\hat{M}}
+ \hf \,\O_{\hat{A}}{}^{\hb\hc}|\,M_{\hb\hc}
+ \F_{\hat{A}}{}^{kl}|\,J_{kl}~.
\eea
The coordinates $x^{\hat m}$ parametrize a curved spacetime $\cM^5$, 
the bosonic body of the superspace $\cM^{5|8}$.

The conditions \eqref{Ddim1} mean that the gravitini can completely be gauged away
such that the projection of the vector covariant derivatives is
\bea
\cD_\ha|=\bD_\ha \quad \Longleftrightarrow \quad \psi_{\hat{m}}{}^\hal_i =0~,
\eea
where 
\bea
\bD_\ha=e_\ha
+\hf\o_\ha{}^{\hb\hc} M_{\hb\hc}
+\phi_\ha{}^{kl} J_{kl},~~~~~~
e_\ha:=e_\ha{}^{\hat{m}} \pa_{\hat{m}}
\eea
is a spacetime covariant derivative with Lorentz and SU(2) connections.
In what follows, we always assume that the gravitini have been gauged away. 
The bosonic covariant derivatives obey  commutation relations of the form
\bea
{[}\bD_\ha,\bD_\hb{]}
&=&
\hf\cR_{\ha\hb}{}^{\hc\hd}M_{\hc\hd}
+\cR_{\ha\hb}{}^{kl}J_{kl}
~,
\eea
where the spacetime curvature tensor $\cR_{\ha\hb}{}^{\hc\hd}$
and the SU(2) field strength $\cR_{\ha\hb}{}^{kl}$ are related to the superspace
 ones as
\bea
\cR_{\ha\hb}{}^{\hc\hd}=R_{\ha\hb}{}^{\hc\hd}|
~,~~~~~~
\cR_{\ha\hb}{}^{kl}
=R_{\ha\hb}{}^{kl}|
~.
\eea

We introduce tensor fields associated with the superspace  dimension-1 torsion tensors:
\bea
&{s}^{kl} := S^{kl}|~, \qquad 
c_\ha{}^{kl} := C_{\ha}{}^{kl}|~, \qquad 
x_{\ha\hb} := X_{\ha\hb} |~, \qquad 
n_{\ha\hb} := N_{\ha\hb} |
~.
\eea
Bar-projecting the super-Weyl tensor gives 
\bea
w_{\ha \hb} := W_{\ha\hb}| = \hf(2 x_{\ha\hb} -n_{\ha\hb})
~.
\label{w4.9}
\eea

\subsection{Conformal Killing spinors}

In this subsection we wish to look for those curved superspace backgrounds which admit 
at least one conformal supersymmetry. 
Such a superspace possesses 
a conformal Killing supervector field $\x^{\hat{A}}$
with the property 
\bea
\x^\ha |=0~, \qquad \e^\hal_i := \x^\hal_i | \neq 0~.
\label{4.100}
\eea
All other bosonic parameters will also be assumed to vanish, 
$K^{\ha \hb}| =0$, $K^{ij}|=0$ and $\s|=0$. 
The spinor parameter  $\e^\hal_i $ generates a $Q$-supersymmetry 
transformation, while  the $S$-supersymmetry transformations 
are generated by 
\bea
\eta^i_\hal := \cD^i_\hal \s|
~.
\eea
With the previous assumptions at hand, bar-projecting the equation \eqref{conf-Killing-0} 
gives\footnote{In what follows, we will sometimes avoid writing 
 spinor indices explicitly. In particular, we will denote  $\e^i:=\e_\hal^i$, and
use   $\G_\ha$ and  $\S_{\ha\hb}$
for $(\G_\ha)_\hal{}^{\hbe}$ and $(\S_{\ha\hb})_\hal{}^\hbe$, 
respectively.
}
\bea
\bD_\ha\e^k
-\hf \Big{[}\,s^k{}_l  \G_{\ha} 
+  \d^k_lx_{\ha\hb}\G^{\hb} 
-\frac{1}{4}    \d^k_l  \ve_{\ha\hb\hc\hd\he}n^{\hb\hc}\S^{\hd\he}
-      c^\hb{}^k{}_l  \S_{\ha\hb} 
\Big{]}
\e^l
-\frac{\ri}{2}\G_\ha\eta^k =0
~,~~~~~~
\label{conf-Killing-spinor}
\eea
which implies
\bea
5 \eta^{i}
&=&
{2\ri} \G^\ha\bD_\ha\e^i
+ \ri \Big{[}\,
{2}  c_\ha{}^i{}_j\G^\ha  + 5 s^{i}{}_{j}{\mathbbm{1}}
+   \d^i{}_j(4x_{\ha\hb}+3  n_{\ha\hb})  \S^{\ha\hb}
\Big{]}
\e^j
~.
\label{conf-Killing-eta}
\eea
The  spinor equation \eqref{conf-Killing-spinor} becomes 
\bea
\bD_\ha\e^{k} =
\frac{1}{2}\S_\ha{}^\hb\bD_\hb\e^{k}
+\frac{1}{8}\Big{(}
3w_{\ha\hb}\G^{\hb}
+\ve_{\ha\hb\hc\hd\he}w^{\hb\hc}\S^{\hd\he}
\Big{)}
\e^k
+\frac{1}{4}\Big{(}c_\ha{}^k{}_{l}\mathbbm{1}
-\frac{1}{2}c^\hb{}^k{}_{l}\S_{\ha \hb}
\Big{)}
\e^l 
~.~~~~~~
\label{conf-Kill-spinor-2}
\eea
This equation may be rewritten in a simpler form if we introduce 
covariant derivatives with torsion, 
\bea
\hat{\bD}_\ha
&:=&
\bD_\ha
-\frac{1}{4}c_\ha{}^{pq}J_{pq}
-\frac{1}{4}\ve_{\ha\hb\hc\hd\he}w^{\hb\hc}M^{\hd\he}
~. 
\label{hatcovder}
\eea
Then \eqref{conf-Kill-spinor-2} turns into 
\bea
\hat{\bD}_\ha\e^k
=
-\frac{1}{5}\G_\ha \G^\hb\,\hat{\bD}_\hb\e^{k}
~.
\label{conf-Kill-spin}
\eea
This is a generalization of the 5D equation for twistor spinors (see, e.g., \cite{deMH1,deMH2}), 
which makes use of the torsion-free covariant derivative $\nabla_\ha$ 
(the Levi-Civit\`a connection) instead of $\hat{\bD}_\ha$.

An important property of twistor spinors is that they `square' to Killing vector fields
\cite{deMH1,deMH2}. This property remains valid in our case. 
Associated with a non-zero {\it commuting} spinor $\e^i$
is the  {\it non-zero} real 5-vector 
\bea
V_\ha
:=
(\G_\ha)^{\hal\hbe}\ve_{ij}\,\e_\hal^{i}\e_\hbe^{j}~.
\label{x-from-e}
\eea
If  $\e^i$  is a solution of \eqref{conf-Kill-spin}, then $V^\ha$
is a conformal Killing vector field, 
\bea
\bD_{(\ha}V_{\hb)}=\frac{1}{5}\eta_{\ha\hb}\bD_{\hc}V^{\hc}
~.
\label{4.18}
\eea
The torsion tensor does not contribute to this relation. 
It is a short calculation  to check that 
\bea
V^\ha V_\ha = -F^2~, \qquad 
F:=\ve^{\hal\hbe}\ve_{ij}\,\e_\hal^{i}\e_{\hbe}^{j}
~.
\eea
Thus, $F$ being real, $V^\ha$ is time-like or null. 
In the spirit of \cite{GG1}, one can construct different bilinears
from a commuting conformal Killing spinor. These bilinears and their properties 
are given in Appendix B.

By construction, we have the identities 
\bea
\d(\cD_\hal^i S^{kl})= 0~,~~~
\d(\cD_\hal^i C_{\ha}{}^{kl})= 0~,~~~
\d(\cD_\hal^i X_{\ha\hb})= 0~,~~~
\d(\cD_\hal^i N_{\ha\hb})= 0~,
\label{4.42}
\eea
which imply that  the conditions \eqref{Ddim1} are superconformal.
Evaluating explicitly the bar-projection of the left-hand sides in \eqref{4.42}, 
non-trivial information may be extracted. 
We derive
{\allowdisplaybreaks\bsubeq\label{dim-2-eqs}
\bea
\cD_\hal^i \cD^{\hbe (k}\cD_{\hbe}^{l)}\s|
&=&
\e^\hbe_j\Big{[}-2\ri {[}\cD_\hal^i,\cD_\hbe^j {]}S^{kl}|
+4\ve^{ij}\bD_{\hal\hbe}s^{kl}
+4 \, \ve^{ij}c_{\hal\hbe}{}^{(k}{}_{p}s^{l)p}
\non\\
&&~~~~\,
+8\big(x_{\hal\hbe}+n_{\hal\hbe}\big)\big(\ve^{k(i}s^{j)l}+\ve^{l(i}s^{j)k}\big)
\Big{]}
\non\\
&&
-12\ri\eta_{\hal j}\big(\ve^{k(i}s^{j)l}+\ve^{l(i}s^{j)k}\big)
+4\ri \eta_\hal^i s^{kl}
~,\\
%%%
( \G_\ha)^{\hga\hde} \cD_\hal^i \cD_{\hga}^{(k}\cD_{\hde}^{l)}\s|
&=&
\e^\hbe_j \Big{[}
-\ri{[}\cD_\hal^i ,\cD_\hbe^j{]}C_{\ha}{}^{kl}|
+2\ve^{ij}\bD_{\hal\hbe}c_{\ha}{}^{kl}
+2\ve_{\hal\hbe}\ve^{ij}x_{\ha\hd}c^{\hd}{}^{kl}
\non\\
&&~~~~
+{1\over 2} \ve^{ij}\ve_{\ha\hb\hc\hd\he}(\G^\hb)_{\hal\hbe}n^{\hc\hd}c^{\he}{}^{kl}
-\ve_{\ha\hb\hc\hd\he}(\S^{\hb\hc})_{\hal\hbe}c^{\hd}{}^{ij}c^{\he}{}^{kl}
\non\\
&&~~~~
-4s^{ij}(\S_{\ha\hd})_{\hal\hbe}c^{\hd}{}^{kl}
-6\ve_{\hal\hbe}\ve^{ij}s^{(k}{}_pc_{\ha}{}^{l)p}
+2\ve^{ij}(\G^\hd)_{\hal\hbe}c_\hd{}^{(k}{}_pc_{\ha}{}^{l)p}
\non\\
&&~~~~
+4\big(x_{\hal\hbe}+n_{\hal\hbe}\big)\big(\ve^{k(i}c_{\ha}{}^{j)l}+\ve^{l(i}c_{\ha}{}^{j)k}\big)
\Big{]}
\non\\
&&
+4\ri\eta^{\hbe i}(\S_{\ha\hb})_{\hbe\hal}c^{\hb}{}^{kl}
-6\ri\eta_{\hal j}\big(
\ve^{k(i}c_{\ha}{}^{j)l}
+\ve^{l(i}c_{\ha}{}^{j)k}
\big)
+2\ri\eta_\hal^i c_{\ha}{}^{kl}
~,~~~~~~~~~~~
%%%
\\
 (\S_{\ha\hb})^{\hbe\hga}\cD_\hal^i \cD_\hbe^k\cD_{\hga k}\s|
 &=&
\e^\hbe_j \Big{[}\,
2\ri{[}\cD_\hal^i ,\cD_\hbe^j{]}X_{\ha\hb}|
-4\ve^{ij}\bD_{\hal\hbe}x_{\ha\hb}
+2\ve^{ij}(\G_\hc)_{\hal\hbe}n_{\hd\he}x_{\hat{f}[\ha}\ve_{\hb]}{}^{\hc\hd\he\hat{f}}
\non\\
&&~~~~
-4(\S_{\hc\hd})_{\hal\hbe}c_{\he}{}^{ij}x_{\hat{f}[\ha}\ve_{\hb]}{}^{\hc\hd\he\hat{f}}
-16s^{ij}(\S_{[\ha}{}^{\hd})_{\hal\hbe}x_{\hb]\hd}
\Big{]}
\non\\
&&
+16\ri\eta^{\hbe i}(\S_{[\ha}{}^{\hc})_{\hbe\hal}x_{\hb]\hc}
-4\ri\eta_\hal^i x_{\ha\hb}
~,
%%%
\\
(\S_{\ha\hb})^{\hbe\hga}\cD_\hal^i \cD_\hbe^k\cD_{\hga k}\s|
&=&
\e^\hbe_j \Big{[}\,
\ri{[}\cD_\hal^i ,\cD_\hbe^j{]}N_{\ha\hb}|
- 2 \,\ve^{ij}\bD_{\hal\hbe}n_{\ha\hb}
+4 \,\ve_{\hal\hbe}\ve^{ij}x_{[\ha}{}^{\hd}n_{\hb]\hd}
\non\\
&&~~~~
+\ve^{ij}(\G_\hc)_{\hal\hbe}n_{\hd\he}n_{\hat{f}[\ha}\ve_{\hb]}{}^{\hc\hd\he\hat{f}}
-2(\S_{\hc\hd})_{\hal\hbe}c_{\he}{}^{ij}n_{\hat{f}[\ha}\ve_{\hb]}{}^{\hc\hd\he\hat{f}}
\non\\
&&~~~~
-8 \,s^{ij}(\S_{[\ha}{}^{\hd})_{\hal\hbe}n_{\hb]\hd}
\Big{]}
\non\\
&&
+8\ri\eta^{\hbe i}(\S_{[\ha}{}^{\hc})_{\hbe\hal}n_{\hb]\hc}
-2\ri\eta_\hal^i n_{\ha\hb}
~.
\eea
\esubeq }
These identities become especially useful for those supersymmetric backgrounds 
which correspond to Poincar\'e or anti-de Sitter supergravities. 

%%%%%%%%%%%%%%%%%%%%%%%%%%%%%%%%%%%%%%%%%

\subsection{Killing spinors}

In the case of Poincar\'e or anti-de Sitter supergravities,
the equations given in the previous subsection 
have to be supplemented by the additional condition 
\bea
\s[\x] =0 \quad \Longrightarrow \quad  \eta^i =0
~,
\label{SzeroEta}
\eea
in accordance with eq. \eqref{constr-sigma}.
Let us remind the reader  that we are not yet  specifying any particular compensators. 
However, we are assuming that some compensator has been chosen 
and the gauge condition \eqref{Xgauge1} has been imposed.

Due to eq.  \eqref{SzeroEta}, the 
 equation for conformal Killing spinors, eq. 
 \eqref{conf-Killing-spinor},  turns into 
\bea
\bD_\ha\e^k
&=&
\Big(
{1\over 2} s^k{}_l\G_\ha
+{1\over 2}  \d_l^k x_{\ha\hb}\G^\hb
-{1\over 8}\d_l^k\ve_{\ha\hb\hc\hd\he}n^{\hb\hc}\S^{\hd\he}
-{1\over 2}c_\hb{}^k{}_l\S_\ha{}^{\hb}
\Big)
\e^l
~.~~~~~~
\label{Killing-spinor3/2}
\eea
Its solutions will be called Killing spinors. 
As demonstrated earlier, associated with a commuting  conformal Killing spinor $\e^k$ is 
the conformal Killing vector $V_\ha$ defined by eq. \eqref{x-from-e}.  
In the case that  $\e^k$ is  a Killing spinor field,  it is simple to prove that
$\bD_\ha V^\ha=0$ and hence 
\bea
\bD_{(\ha}V_{\hb)}=0
~.
\label{4.24}
\eea
Thus $V_\ha$ is  a Killing vector field.

%%%%%%%%%%%%%%%%%%%%%%%%%%%%%%%%%%%%%%%%%%%%%%%%%%%%%%

\section{Supersymmetric backgrounds: Eight supercharges}
\label{8-susy}
\setcounter{equation}{0}

The existence of rigid supersymmetries imposes non-trivial restrictions on 
the background fields in off-shell Poincar\'e or anti-de Sitter supergravities.  
For simplicity, here we  restrict our analysis to the case of eight supercharges and derive 
constraints on the geometry.

Since $\s[\x] =0$, the equations \eqref{dim-2-eqs}  
immediately imply the following  conditions:
\bea
{[}\cD_\hal^i,\cD_\hbe^j {]}S^{kl}|=0~,~~
{[}\cD_\hal^i ,\cD_\hbe^j{]}C_{\ha}{}^{kl}|=0
~,~~
{[}\cD_\hal^i ,\cD_\hbe^j{]}X_{\ha\hb}|=0
~,~~
{[}\cD_\hal^i ,\cD_\hbe^j{]}N_{\ha\hb}|=0
~.
~~~~~~~~~
\label{dim-2-constr}
\eea
The meaning of these conditions is  that all dimension-2 auxiliary fields, 
which belong to the supergravity multiplet, vanish.   
Information about the background geometry is encoded
in the background dimension-1 fields
 $s^{ij}$, $c_\ha^{ij}$, $x_{\ha\hb}$ and $n_{\ha\hb}$.
The same equations \eqref{dim-2-eqs} 
also lead to  a set of conditions on these  tensors.
Below we describe the various cases by the values of 
$s:= \sqrt{\hf s^{ij}s_{ij}}$ and $c_\ha{}^{ij}$.

The relations \eqref{dim-2-constr} are in fact corollaries of more general results 
that follow from the following observation.  
For any background admitting eight supercharges, 
if there is a  tensor superfield $T$ such that 
its bar-projection vanishes, $T|=0$, and this condition is supersymmetric,
then the entire superfield is zero, $T=0$.
For all supersymmetric backgrounds, the conditions
\eqref{Ddim1} hold. Therefore, all backgrounds with eight supercharges 
should fulfil the superfield conditions
\bea
\cD_\hal^i S^{kl}= 0~, \qquad 
\cD_\hal^i C_{\ha}{}^{kl}= 0~, \qquad 
\cD_\hal^i X_{\ha\hb}= 0~, \qquad 
\cD_\hal^i N_{\ha\hb}= 0~. 
\eea
The relations \eqref{dim-2-constr} obviously follow from these conditions.

\subsection{The case $s\ne 0$} \label{section5.1}

When $s\ne 0$, it can be shown that eqs. \eqref{dim-2-eqs} imply
the conditions:
\bsubeq \label{constr-s}
\bea
&\bD_{\ha}s^{ij}=0\quad \Longrightarrow \quad
s={\rm const}
~,
\label{constr-s-1}
\\
&c_\ha{}^{ij}=0
~,~~~~~~
x_{\ha\hb}=0
~,~~~~~~
n_{\ha\hb}=0
~.
\label{constr-s-2}
\eea
\esubeq
The Killing spinor equation takes the simple form
\bea
\bD_\ha\e^k
=
{1\over 2} s^k{}_l\G_\ha \e^l
~.
\label{Killing-s}
\eea

By computing  ${[}\bD_{\ha},\bD_{\hb}{]}\e^k$ and using \eqref{Killing-s} together with 
\eqref{constr-s} one  obtains
\bea
{[}\bD_\ha,\bD_\hb{]}\e^k
=
-s^2\S_{\ha\hb} \e^k
=
\Big(
\hf\d^k_l\cR_{\ha\hb}{}^{\hc\hd}\S_{\hc\hd}
+\cR_{\ha\hb}{}^k{}_{l}\mathbbm{1}
\Big)\e^l
~,
\eea
from which we can read off the expressions for the Lorentz and SU(2) curvatures\footnote{Note that here
 we compute the component curvature tensors by using the 
 Killing spinor equation \eqref{Killing-s}. 
 The same results can be read off
 by bar-projecting the dimension-2 superspace curvature tensors.}
\bsubeq\label{sn0curvatures}
\bea
\cR_{\ha\hb}{}^{\hc\hd}
&=&
-2s^2\d_{[\ha}^\hc\d_{\hb]}^\hd
~,
\\
\cR_{\ha\hb}{}^{kl}
&=&0
~.
\eea
\esubeq
Hence in this case the supersymmetric background is necessarily 
5D anti-de Sitter space, AdS$_5$. It follows from \eqref{constr-s-2} that 
three  dimension-1 superfield torsion tensors vanish, 
\bea
C_\ha{}^{ij}=0
~,~~~~~~
X_{\ha\hb}=0
~,~~~~~~
N_{\ha\hb}=0
~.
\eea
The resulting superspace AdS$^{5|8} $ and rigid supersymmetric field theories 
in AdS$^{5|8} $ have thoroughly  been studied in \cite{KT-M07,KT-M5DCF}.

\subsection{The case $s=0$ and $c_\ha{}^{ij}\ne0$} \label{section5.2}

If  $s=0$ and $c_\ha{}^{ij}\ne0$,  
the relations \eqref{dim-2-eqs} imply that
some of the background fields vanish, 
\bea
s^{ij}=0
~,~~~~~~
x_{\ha\hb}=0
~,~~~~~~
n_{\ha\hb}=0~, \label{5.8}
\eea
as well as  the following constraints on $c_\ha{}^{ij}$
\bea
\bD_\ha c_{\hb}{}^{ij}=0
~,~~~~~~
c_{\ha}{}^{(i}{}_k   c_{\hb}{}^{j)k}=0
~.
\eea
These constraints tell us that 
$c_\ha{}^{ij}$ is a composite object being 
the product of a real 5-vector $c_\ha$ 
and a real  isovector $c^{ij}$ such that  $\overline{c^{ij}}= c_{ij}$, 
\bea
c_\ha{}^{ij}=c_\ha c^{ij}
~.
\label{constr-c-1}
\eea
By rescaling $c_\ha$ and $c^{ij}$ we can always make the choice
\bea
c^{ij}c_{ij}= 2
~.
\label{constr-c-3}
\eea
Then it follows from $\bD_\ha c_{\hb}{}^{ij}=0$ that 
 $c_\ha$ 
and  $c^{ij}$ are covariantly constant,
\bea
\bD_\ha c^{ij}=0~,~~~
\bD_\ha c_\hb=0
~.
\label{constr-c-2}
\eea
The 5-vector $c^\ha$ may be time-like, space-like or null. 
Since it is covariantly constant, 
the Lorentz  curvature tensor is constrained by 
\bea
\cR_{\ha\hb \hc\hd} \,c^\hd = 0~.
\label{5.12}
\eea

For the background under consideration, the Killing spinor equation is
\bea
\bD_\ha\e^k
&=&
-{1\over 2} c^\hb c^k{}_l\S_{\ha\hb}\e^l
~.
\label{Killing-c}
\eea
We compute  ${[}\bD_{\ha},\bD_{\hb}{]}\e^k$  
by making use of \eqref{Killing-c} in conjunction 
with eqs. \eqref{constr-c-1} and \eqref{constr-c-3}. 
The result is
\bea
{[}\bD_\ha,\bD_\hb{]}\e^k
&=&
-\frac{1}{2}\Big(
c_{[\ha} \d_{\hb]}^{[\hc} c^{\hd]} 
+\hf(c^\he c_\he)\d_{[\ha}^\hc\d_{\hb]}^\hd
\Big)\S_{\hc\hd}
\e^k
~,
\eea
from which we can read off the Lorentz and SU(2) curvature tensors
\bsubeq
\bea
\cR_{\ha\hb}{}^{\hc\hd}&=&
-\Big(
c_{[\ha} \d_{\hb]}^{[\hc} c^{\hd]} 
+\hf(c^\he c_\he)\d_{[\ha}^\hc\d_{\hb]}^\hd
\Big)
~, \label{5.15a}
%%%%%%%%%%%%
\\
\cR_{\ha\hb}{}^{kl}&=&0
~.
\eea
\esubeq
It follows from  \eqref{5.15a} that 
the Ricci tensor and the scalar curvature are
\bea
\cR_{\ha\hb}
=
\frac{3}{4}\Big(
c_{\ha} c_{\hb} 
-\eta_{\ha\hb} c^{\he} c_{\he}
\Big)
~,~~~~~~
\cR
=
-3c^{\he} c_{\he}~.
\label{5.16}
\eea
As concerns the Weyl tensor 
\bea
C_{\ha\hb\hc\hd}
&=&
\cR_{\ha\hb\hc\hd}
-\frac{2}{3}\Big(\eta_{\ha[\hc}\cR_{\hd]\hb}-\eta_{\hb[\hc}\cR_{\hd]\ha}\Big)
+\frac{1}{6}\cR\eta_{\ha[\hc}\eta_{\hd]\hb} ~,
\eea
it is identically zero for the above background, 
\bea
C_{\ha\hb\hc\hd} =0~.
\label{5.18}
\eea
The above supersymmetric backgrounds are generalizations of those 
found by Festuccia and Seiberg \cite{FS} in the case of the old 
minimal formulation for 4D $\cN=1$ supergravity.

The existence of a covariantly constant vector field 
$c^\ha$ means that spacetime is decomposable
in the non-null case  (see, e.g., \cite{Stephani}).
In this case 
the space is the product of a four- and a one-dimensional manifold. 
We can choose a coordinate frame $x^{\hat m} = (x^m, \z) $, 
where $m =1,2,3,4$, 
such that the vector field $ c^\ha e_\ha$ is proportional to $\pa/ \pa \z$
and the metric reads
\bea
\rd s_5^2 = g_{{m} {n} } (x^r)  \rd x^m \rd x^n +\ve (\rd \z)^2 
= \eta_{ab} e^a e^b +\ve (\rd \z)^2~, \qquad e^a:=\rd x^m e_m{}^a (x^n)~,
\eea 
where $\ve =-1$  when $c^\ha$ is time-like, and $\ve =+1$ when $c^\ha$ is space-like. 
The metric $\rd s_4^2 = g_{{m} {n} } (x^r)  \rd x^m \rd x^n $ corresponds to a 
four-dimensional submanifold $\cM^4$ orthogonal to $ c^\ha e_\ha$. 
The identity \eqref{5.12} tells us 
\bea
\cR_{\ha\hb \hc  \,\z } =0~.
\eea
Let $\frak R_{ab cd}$ be the curvature of the submanifold $\cM^4$.
It is clear that 
\bea
\frak R_{a b c  d } = \cR_{a b c  d } ~.
\eea
Then from \eqref{5.16} and \eqref{5.18} we deduce
\bea
{\frak R}_{ab} = -\frac{3}{4} c^2 \eta_{ab} ~, \qquad {\frak C}_{abcd} =0~,
\eea
where $c^2 = c^{\he} c_{\he}$. We conclude that $\cM^4 $ is a four-sphere, 
$S^4$, when $c^\ha$ is time-like. 
In the case that  $c^\ha $ is space-like, $\cM^4$ is a four-dimensional anti-de Sitter space, AdS$_4$. 

Finally, if $c^\ha$ is  null, $c^\ha c_\ha=0$,
it is possible to chose a coordinate system  in which
the metric reads 
\bea
\rd s^2 = \re^u (2\rd u\, \rd v + \d_{ij} \rd x^i \rd x^j) ~, \qquad i,j =1,2,3~,
\eea 
with $c^\ha e_\ha \propto \pa/ \pa v$.
This is a special example of pp-waves, see, e.g., \cite{Ortin}.

In conclusion, we present those superspace geometries that generate
the supersymmetric backgrounds given. It follows from \eqref{5.8} 
that $ S^{ij}=0$,  $X_{\ha\hb}=0$ and $N_{\ha\hb}=0$.
The superspace geometry is described  
by a single  covariantly constant tensor $C_\ha{}^{ij}$, $\cD_{\hat{B}}C_\ha{}^{ij}=0$.
The algebra of covariant derivatives is
\begin{subequations} \label{5.24}
\bea
\big\{ \cD_{\hal}^i , \cD_{\hbe}^j \big\} &=&-2 \ri \,\ve^{ij}\cD_{\hal\hbe}
-{\ri\over 2}\ve^{\ha\hb\hc\hd\he}(\S_{\ha\hb})_{\hal\hbe}C_{\hc}{}^{ij}M_{\hd\he}
-\ri \, \ve^{ij}C_{\hal\hbe}{}^{kl}J_{kl}
~,
 \\
%%%%%%
{[}\cD_\ha,\cD_{\hbe}^j{]}&=&
{1\over 2}(\S_\ha{}^{\hb})_{\hbe}{}^{\hga}C_\hb{}^j{}_k\cD_{\hga}^k
~,
\\
%%%%%%
{[}\cD_\ha,\cD_\hb{]}
&=&
\frac{1}{4} 
\Big(\d_{[\ha}^{[\hc} C_{\hb ] kl} C^{\hd] kl} 
- \frac{1}{2} \d_{[\ha}^\hc \d_{\hb]}^\hd C^{\he kl} C_{\he kl} 
\Big)
M_{\hc\hd}
~.
\eea
\end{subequations}
Integrability condition for the constraint $\cD_\hal^i C_{\hb}{}^{jk}=0$ is
\bea
C_\ha{}^{(i}{}_kC_\hb{}^{j)k}=0
~.
\eea
It implies that the superfield $C_\ha{}^{ij}$ factorizes, 
\bea
C_\ha{}^{ij}=C_\ha C^{ij}
~,~~~~~~
C^{ij}C_{ij}=2
~.~~~
\eea
The condition that $C_{\ha}{}^{ij}$ is covariantly constant is equivalent to 
$
C_{ij}\cD_{\hat{A}} C_\hb = -  C_\hb\cD_{\hat{A}} C_{ij}$,
which leads to
\bea
2\cD_{\hat{A}} C_\hb =  -  C_\hb C^{ij}\cD_{\hat{A}} C_{ij} = -  \hf C_\hb \cD_{\hat{A}}(C^{ij} C_{ij}) = 0
~.
\eea
Thus both tensors $C_\ha$ and $C^{ij}$ are covariantly constant, 
\bea
\cD_{\hat{A}} C_\hb=0~, \qquad \cD_{\hat{A}} C^{ij}=0~.
\eea
Because the superspace is conformally flat, the isometry superalgebra is a subalgebra 
of the 5D superconformal algebra ${\frak f}(4)$.

\subsection{The case $s = 0$ and $c_\ha{}^{ij} = 0$} \label{section5.3}

It remains to consider the case
\bea
s^{ij}=0~,~~~~~~
c_\ha{}^{ij}=0~.
\label{constr-x-0}
\eea
Here  the relations \eqref{dim-2-eqs} imply 
 the following constraints on $x_{\ha\hb}$ and $n_{\ha\hb}$
\bsubeq \label{constr-x}
\bea
\bD_\ha x_{\hb\hc}&=&
\hf \ve_{\ha\hd\he\hat{f}[\hb} x_{\hc]}{}^{\hd}n^{\he\hat{f}}
~,
\label{constr-x-1}
\\
\bD_\ha n_{\hb\hc}&=&
\hf \ve_{\ha\hd\he\hat{f}[\hb}n_{\hc]}{}^{\hd}n^{\he\hat{f}}
=-\frac{1}{8} \eta_{\ha[\hb}\ve_{\hb]\hd\he\hat{f}\hat{g}}n^{\hd\he}n^{\hat{f}\hat{g}}
~,
\label{constr-x-2}
%%%
\\
x_{[\ha}{}^{\hc}n_{\hb]\hc}
&=&0
~.~~~~~~
\label{constr-x-3}
\eea
\esubeq
The constraint \eqref{constr-x-3} can be rewritten in a matrix form 
as follows:
\bea
[\hat{x},\hat{n}]=0   ~, \qquad \hat{x}: = (x_\ha{}^\hb) ~, \quad 
\hat{n}:=(n_\ha{}^\hb)~.
\eea

An important consequence of the constraints \eqref{constr-x-1} and
\eqref{constr-x-3} is 
that $x_{\ha\hb}$ is a closed two-form,
\bea
\bD_{[\ha}x_{\hb\hc]}=0
~.
\eea
It is a consequence of \eqref{constr-x-2} that $n_{\ha\hb}$ 
is  also a closed two-form,
\bea
\bD_{[\ha}n_{\hb\hc]}=0
~.
\eea
Introducing the Hodge dual of $n_{\ha\hb}$ in the standard way
${}*\!n_{\ha\hb\hc}:=\frac{1}{2}\ve_{\ha\hb\hc\hd\he}n^{\hd\he}$,
the constraint \eqref{constr-x-2} becomes
\bea
\bD_\ha *\!n_{\hb\hc\hd}&=&
-\frac{3}{2}n_{[\ha\hb}n_{\hc\hd]}
~.
\eea
This relation implies the equation of motion that is derived from a U(1) Chern-Simons action. 

For the background under consideration, 
the Killing spinor equation \eqref{Killing-spinor3/2}  takes the form 
\bea
\bD_\ha\e^k
&=&
\Big(
\,
{1\over 2}\d^k_l  x_{\ha\hb}\G^\hb
-{1\over 8}\d^k_l\ve_{\ha\hb\hc\hd\he}n^{\hb\hc}\S^{\hd\he}
\Big)
\e^l
~.
\label{Killing-x}
\eea
We can now compute ${[}\bD_{\ha},\bD_{\hb}{]}\e^k$ 
by using \eqref{Killing-x} together with the relations
\eqref{constr-x-1}--\eqref{constr-x-3}. The result is
\bea
{[}\bD_\ha,\bD_\hb{]}\e^k
&=&
\Big(
-\frac{3}{4}n_{[\ha}{}^{[\hc} n_{\hb]}{}^{\hd]}
+\frac{1}{4}n_{\ha\hb}n^{\hc\hd}
+{1\over 2}\d^{[\hc}_{[\ha}n_{\hb]\he}n^{\hd]\he}
-{1\over 8}n_{\he\hat{f}}n^{\he\hat{f}}\d^{[\hc}_{[\ha}\d^{\hd]}_{\hb]}
\non\\
&&~~
+x_{[\ha}{}^{[\hc}x_{\hb]}{}^{\hd]}
\Big)\S_{\hc\hd}\e^k
~.~~~~~~~~~
\eea
${}$From here 
we read off the Lorentz and SU(2) curvature tensors
\bsubeq
\bea
\cR_{\ha\hb}{}^{\hc\hd}
&=&
2x_{[\ha}{}^{[\hc}x_{\hb]}{}^{\hd]}
+\frac{1}{2}n_{\ha\hb}n^{\hc\hd}
-\frac{3}{2}n_{[\ha}{}^{[\hc} n_{\hb]}{}^{\hd]}
+\d^{[\hc}_{[\ha}n_{\hb]\he}n^{\hd]\he}
-{1\over 4}n_{\he\hat{f}}n^{\he\hat{f}}\d^{[\hc}_{[\ha}\d^{\hd]}_{\hb]}
~,
\label{Lorentz-x-0}
\\
\cR_{\ha\hb}{}^{kl}
&=&
0
~.
\eea
\esubeq

Actually there is another important constraint on 
the dimension-1 tensors $x_{\ha\hb}$ and $n_{\ha\hb}$. 
For the background under consideration, 
it can be proved that the dimension-2 superspace Bianchi identities
imply the following quadratic equation
\bea
x_{[\ha\hb}x_{\hc]\hd}
=
n_{[\ha\hb}n_{\hc]\hd}
~~~~~~
\Longleftrightarrow
~~~~~~
x_{[\ha\hb}x_{\hc\hd]}
=
n_{[\ha\hb}n_{\hc\hd]}
~.
\label{constr-x-4}
\eea
This constraint may be seen to be
equivalent to the requirement  that the Lorentz curvature \eqref{Lorentz-x-0} 
satisfies the Bianchi identity $\cR_{[\ha\hb\hc]\hd}=0$.
With the aid of 
\eqref{constr-x-4} we can rewrite the Lorentz curvature in the equivalent form:
\bea
\cR_{\ha\hb}{}^{\hc\hd}
&=&
-\frac{1}{6}n_{\ha\hb}n^{\hc\hd}
-\frac{1}{6}n_{[\ha}{}^{[\hc} n_{\hb]}{}^{\hd]}
+\d^{[\hc}_{[\ha}n_{\hb]\he}n^{\hd]\he}
-{1\over 4}n_{\he\hat{f}}n^{\he\hat{f}}\d^{[\hc}_{[\ha}\d_{\hb]}^{\hd]}
\non\\
&&
+\frac{2}{3}x_{\ha\hb}x^{\hc\hd}
+\frac{2}{3}x_{[\ha}{}^{[\hc}x_{\hb]}{}^{\hd]}
~.
\eea
The Ricci tensor and the scalar curvature are, respectively,
\bsubeq
\bea
\cR_{\ha\hb}
&=&
x_{\ha}{}^{\hc}x_{\hb\hc}
+\frac{1}{2}n_{\ha}{}^{\hc}n_{\hb\hc}
-\frac{1}{4}\eta_{\ha\hb}\, n^{\hc\hd}n_{\hc\hd}
~,
%%%%%%
\\
\cR
&=&
x^{\ha\hb}x_{\ha\hb}
-\frac{3}{4}n^{\ha\hb}n_{\ha\hb}
~.
\eea
\esubeq
The Weyl tensor is
\bea
C_{\ha\hb\hc\hd}
&=&
-\frac{1}{6}\Big(
n_{\ha\hb}n_{\hc\hd}
-n_{\hc[\ha} n_{\hb]\hd}
-\eta_{\hc[\ha}n_{\hb]}{}^{\he}n_{\hd\he}
+\eta_{\hd[\ha}n_{\hb]}{}^{\he}n_{\hc\he}
+\frac{1}{4}\eta_{\ha[\hc}\eta_{\hd]\hb}n^{\he\hat{f}}n_{\he\hat{f}}
\Big)
\non\\
&&
+\frac{2}{3}\Big(
x_{\ha\hb}x_{\hc\hd}
-x_{\hc[\ha}x_{\hb]\hd}
-\eta_{\hc[\ha}x_{\hb]}{}^{\he}x_{\hd\he}
+\eta_{\hd[\ha}x_{\hb]}{}^{\he}x_{\hc\he}
+\frac{1}{4}\eta_{\ha[\hc}\eta_{\hd]\hb}x^{\he\hat{f}}x_{\he\hat{f}}
\Big)
~.~~~~~~~~~
\label{Weyl-components}
\eea

An important observation is in order. 
It may be seen that the Weyl tensor \eqref{Weyl-components}  vanishes
(and the spacetime is conformally flat), $C_{\ha\hb\hc\hd}=0$,
under the condition
\bea
w_{\ha\hb}:=W_{\ha\hb}| =  0~~~~~~
\Longleftrightarrow
~~~~~~
n_{\ha\hb}=2x_{\ha\hb}
~.
\eea
Due to \eqref{constr-x-4}, in this case we should also have the condition
$x_{[\ha\hb}x_{\hc\hd]}=0$, which is equivalent to 
the fact  that $x_{\ha\hb}$ is a decomposable 
bivector, $x_{\ha\hb} =u_{[\ha} v_{\hb]}$, for some 5-vectors
$  u_\ha $ and $v_\ha$. 
Then we deduce from \eqref{constr-x} 
that the  two-form $x_{\hb\hc}$ is covariantly constant, 
\bea
\bD_{\ha}x_{\hb\hc}=0~.
\eea

We now present the superspace geometry that generates the 
bosonic background described. In accordance with \eqref{constr-x-0}, 
the dimension-1 torsion tensors $S^{ij}$ and $C_\ha{}^{ij}$ vanish. 
The superspace geometry is determined 
by the tensors
$X_{\ha\hb}$ and  $N_{\ha\hb}$ 
obeying the differential constraints 
\bsubeq 
\bea
\cD_\hal^i X_{\ha\hb}&=&0
~,~~~~~~
\cD_\hal^i N_{\ha\hb}=0~, 
\\
\cD_\ha X_{\hb\hc}&=&
\hf \ve_{\ha\hd\he\hat{f}[\hb} X_{\hc]}{}^{\hd}N^{\he\hat{f}}
~,
\\
\cD_\ha N_{\hb\hc}&=&
\hf \ve_{\ha\hd\he\hat{f}[\hb}N_{\hc]}{}^{\hd}N^{\he\hat{f}}
=-\frac{1}{8} \eta_{\ha[\hb}\ve_{\hc]\he\hd\hat{f}\hat{g}}N^{\hd\he}N^{\hat{f}\hat{g}}
\eea
and the algebraic ones
\bea
X_{[\ha}{}^{\hc}N_{\hb]\hc}
&=&0
~,~~~~~~
X_{[\ha\hb}X_{\hc]\hd}
=
N_{[\ha\hb}N_{\hc]\hd}
~.
\eea
\esubeq
The  algebra of covariant derivatives is
\begin{subequations} \label{5.46}
\bea
\big\{ \cD_{\hal}^i , \cD_{\hbe}^j \big\} &=&-2 \ri \,\ve^{ij}\cD_{\hal\hbe}
-\ri \,\ve_{\hal\hbe}\ve^{ij}X^{\hc\hd}M_{\hc\hd}
+{\ri\over 4} \ve^{ij}\ve^{\ha\hb\hc\hd\he}(\G_\ha)_{\hal\hbe}N_{\hb\hc}M_{\hd\he}
\non\\
&&
-4\ri\Big(X_{\hal\hbe}+N_{\hal\hbe}\Big)J^{ij}
~,
%%%%%%
 \\
{[}\cD_\ha,\cD_{\hbe}^j{]}&=&
-{1\over 2} \Big(
 X_{\ha\hb}(\Gamma^{\hat{b}})_{\hbe}{}^{\hga}
+{1\over 4}\,\ve_{\ha\hb\hc\hd\he}N^{\hd\he}(\Sigma^{\hb\hc})_{\hbe}{}^{\hga}
\Big)
\cD_{\hga}^j
~,
%%%%%%
\\
{[}\cD_\ha,\cD_{\hb}{]}
&=&
-\hf \Big(\,
\frac{1}{6}N_{\ha\hb}N^{\hc\hd}
+\frac{1}{6}N_{[\ha}{}^{[\hc} N_{\hb]}{}^{\hd]}
-\d^{[\hc}_{[\ha}N_{\hb]\he}N^{\hd]\he}
+{1\over 4}N_{\he\hat{f}}N^{\he\hat{f}}\d^{[\hc}_{[\ha}\d_{\hb]}^{\hd]}
\non\\
&&~~~~~
-\frac{2}{3}X_{\ha\hb}X^{\hc\hd}
-\frac{2}{3}X_{[\ha}{}^{[\hc}X_{\hb]}{}^{\hd]}
\Big)
M_{\hc\hd}
~.
\eea
\end{subequations}
This superspace is  conformally flat only if 
$W_{\ha \hb} = X_{\ha\hb} -\hf  N_{\ha\hb} =0$, and then the bivector 
$X_{\ha \hb}$ is covariantly constant and decomposable, 
\bea
W_{\ha\hb}=0 \quad \Longrightarrow \quad \cD_{\hat A} X_{\ha \hb} =0~, 
\qquad
X_{[\ha\hb}X_{\hc]\hd} =0~.
\eea

%%%%%%%%%%%%%%%%%%%%%%%%%%%%%%%%%%%%%%%%%%%%%%%%%%%%%%%%

\section{Vector multiplet compensator}
\setcounter{equation}{0}

Up to now we have not specified any conformal compensator. 
Similar to the case of 4D $\cN=2$ supergravity reviewed in \cite{FVP}, 
two conformal compensators are required in 5D minimal supergravity. 
One of them is universally a vector multiplet, while there are several
choices  for the second compensator. It may be  an $\cO(2)$ multiplet, or 
 a hypermultiplet, or a nonlinear multiplet.
 The dilaton Weyl multiplet automatically includes one compensator, an on-shell 
 vector multiplet.  
In the remainder of this paper, we will study the restrictions on supersymmetric backgrounds
which arise when one or two compensators are turned on.

As mentioned above, one of the compensators is invariably an Abelian vector multiplet.
The standard way to formulate it is to use gauge 
 covariant derivatives
\bea
{\bm \cD}_{\hat{A}}=
\cD_{\hat{A}}
+\ri \cV_{\hat{A}} {\frak Z}~,
\label{6.1}
\eea
where $\frak Z$ denotes the U(1) generator 
and 
$\cV_{\hat{A}}$ is the corresponding  connection.
 In general the gauge covariant derivatives 
have (anti-)commutation relations 
\bea
{[}{\bm \cD}_{\hat{A}},{\bm \cD}_{\hat{B}}\}&=&
T_{\hat{A}\hat{B}}{}^{\hat{C}}\, {\bm \cD}_{\hat{C}}
+\hf R_{\hat{A}\hat{B}}{}^{\hc\hd}M_{\hc\hd}
+R_{\hat{A}\hat{B}}{}^{kl}J_{kl}
+\ri F_{\hat{A}\hat{B}}\frak Z~,
\eea
in which the torsion, and the Lorentz and SU(2) curvature tensors are the same as before.
In order to describe the vector multiplet, 
the U(1) field strength $F_{\hat{A}\hat{B}}$ 
is constrained such that its components are \cite{KT-Mconfsugra5D}
\begin{subequations}\label{FS6.3}
\bea
&&F_\hal^i{}_\hbe^j=-2\ri\ve^{ij}\ve_{\hal\hbe}W~,\qquad 
F_{\ha}{}_\hbe^j=(\G_\ha)_\hbe{}^{\hga}\cD_{\hga}^jW~,
\label{FSa}\\
&&F_{\ha\hb}=
X_{\ha\hb}W
+{\ri\over 4}(\S_{\ha\hb})^{\hga\hde}\cD_{\hga}^k\cD_{\hde k}W~.
\label{FSb}
\eea
\end{subequations}
Here the field strength $W$  is real, $\bar W = W$, and obeys the Bianchi identity
\bea
\cD_{\hal}^{(i}\cD_{\hbe}^{j)}W
-{1\over 4}\ve_{\hal\hbe}\cD^{\hga(i}\cD_{\hga}^{j)} W 
= {\ri\over 2} C_{\hal\hbe}{}^{ij} W ~.~~~~~~
\label{W-BI}
\eea
The super-Weyl transformation law of the field strength $W$ is
\bea
\d_\s W=\s W~.
\label{super-Weyl-W}
\eea

We require the field strength  $W$ to be  nowhere vanishing,
$W >0$, so that it can be used as a conformal compensator.
Actually, since $W $ is a Lorentz and SU(2) scalar superfield,  
it can be identified with the compensating 
superfield $\F$ introduced in section \ref{Isometries}.
Choosing the super-Weyl gauge
\bea
W =1~,
\label{minimal-1}
\eea
completely fixes the super-Weyl gauge freedom. 
This gauge choice leads to 
 the following restrictions on the dimension-1 torsion superfields:
\bea
X_{\ha\hb}=F_{\ha\hb}
~,~~~~~~
C_{\ha}{}^{kl} = 0 ~.
\label{minimal-2}
\eea

The superspace geometry described by the gauge covariant derivatives 
${\bm \cD}_{\hat{A}}$ and subject to the condition \eqref{minimal-1} 
corresponds to the 5D $\cN=1$ minimal  supergravity multiplet. 
It was discovered by Howe \cite{Howe5Dsugra} in 1982
in the superspace setting and then was fully elaborated in 
\cite{KT-Msugra5D,KT-Msugra5D2}. In the component approach, 
the minimal multiplet was rediscovered by Zucker in 
1999 \cite{Zucker}.

\subsection{Supersymmetric backgrounds}

All information about the supersymmetric backgrounds that 
correspond to the minimal supergravity multiplet can be extracted from the results 
in sections 4 and 5. It suffices to take into account the conditions
\eqref{minimal-1} and \eqref{minimal-2}.
In particular, the Killing spinor equation \eqref{Killing-spinor3/2} turns into 
\bea
\bD_\ha\e^k
&=&
\Big(
{1\over 2} \G_\ha s^k{}_l
+{1\over 2}\d_l^k  f_{\ha\hb}\G^\hb
-{1\over 8}\d_l^k\ve_{\ha\hb\hc\hd\he}n^{\hb\hc}\S^{\hd\he}
\Big)
\e^l~,
\label{6.8}
\eea
where we have denoted
$f_{\ha\hb}:=F_{\ha\hb}|=x_{\ha\hb}$. 
By construction,  the two-form $f_{\ha\hb}$ is a U(1) field strength,  
$\bD_{[\ha}f_{\hb\hc]}=0$.

All supersymmetric backgrounds with eight supercharges 
are characterized by the conditions
\bea
f_{\ha\hb}s^{ij}&=&0~,\qquad
n_{\ha\hb}s^{ij}=0~.
\eea
The background fields obey the following differential and algebraic conditions:
\bsubeq
\bea
\bD_{\ha}s^{kl}&=&0
~,\\
\bD_\ha f_{\hb\hc}&=&
\hf \ve_{\ha\hd\he\hat{f}[\hb} f_{\hc]}{}^{\hd}n^{\he\hat{f}}
~,
%%%
\\
\bD_\ha n_{\hb\hc}&=&
\hf \ve_{\ha\hd\he\hat{f}[\hb}n_{\hc]}{}^{\hd}n^{\he\hat{f}}
=-\frac{1}{8} \eta_{\ha[\hb}\ve_{\hc]\hd\he\hat{f}\hat{g}}n^{\hd\he}n^{\hat{f}\hat{g}}
~,
%%%%%%
\\
%%%
f_{[\ha}{}^{\hc}n_{\hb]\hc}
&=&
0
~,~~~~~~
\\
%%%
f_{[\ha\hb}f_{\hc]\hd}
&=&
n_{[\ha\hb}n_{\hc]\hd}
~~~~~~
\Longleftrightarrow
~~~~~~
f_{[\ha\hb}f_{\hc\hd]}
=
n_{[\ha\hb}n_{\hc\hd]}
~.
\eea
\esubeq
The curvature tensors 
can be read off from the results of 
the previous section by setting  $c_\ha{}^{ij}=0$
and $x_{\ha\hb}=f_{\ha\hb}$.

\subsection{The dilaton Weyl multiplet}

In the superspace setting of \cite{KT-Mconfsugra5D}, 
the so-called dilaton Weyl multiplet \cite{Ohashi,Bergshoeff} 
is  realized as the   Weyl
multiplet coupled to an Abelian vector multiplet  such that 
its  field strength $W$ is nowhere vanishing, 
$W  \neq 0$, and enjoys the equation
\bea
{\mathbb H}^{ij} = 0~, 
\label{Chern-Simons}
\eea
where ${\mathbb H}^{ij} $ denotes the following real isovector 
\cite{KT-Mconfsugra5D}
\bea
{\mathbb H}^{ij} &=&\ri\,\cD^{\hal (i}W\cD_\hal^{j)} W+{\ri\over 2} W \cD^{ij} W-2S^{ij}W^2~
=\frac{\ri}{6 W}\big(\cD^{ij}+12\ri S^{ij}\big)W^3~,
\label{mbbH-def}
\eea
which is
 constrained by
\bea
\cD_\hal^{(i} {\mathbb H}^{jk)} &=&0~.
\eea
This constraint defines an $\cO (2) $ multiplet.\footnote{In the rigid supersymmetric case, 
the composite $\cO (2) $ multiplet \eqref{mbbH-def}
was introduced in  \cite{KL}.}  
The super-Weyl transformation law of $\mathbb H^{ij}$ is 
\bea
\d_\s \mathbb H^{ij}= 3\s \mathbb H^{ij}~.
\eea

Eq. \eqref{Chern-Simons} is equivalent to
\be
S^{ij} = \frac{\rm i}{2W^2} \Big\{  
\cD^{\hal (i}W\cD_\hal^{j)} W+\hf W \cD^{ij} W\Big\}~.
\ee
Similar to the rigid supersymmetric case \cite{KL}, 
eq. (\ref{Chern-Simons}) originates as the equation of motion 
in a Chern-Simons model for the vector multiplet.

In the super-Weyl gauge \eqref{minimal-1}, 
we have the condition
\bea
S^{ij}=0~,
\eea
in addition to the superfield requirements \eqref{minimal-2}.

%%%%%%%%%%%%%%%%%%%%%%%%%%%%%%%%%%%%%%%%%%%%%%%%%%

\section{$\cO(2)$ multiplet compensator}\label{section7}
\setcounter{equation}{0}

There are several ways to choose the second supergravity
compensator. Similar to the situation in 4D $\cN=2$ supergravity 
(see, e.g., \cite{FVP} for a review), 
one of the most popular choices is 
a real  $\cO(2)$ multiplet.\footnote{It is a 5D analogue 
of the 4D $\cN=2$ improved tensor multiplet \cite{deWPV,LR83}.} 
Within the superspace approach of \cite{KT-M07},
this multiplet is described by an isovector superfield $H^{ij} = H^{ji} = \ve^{ik}\ve^{jl}
\overline{H^{kl}} $ which is constrained by 
\bea
\cD_\hal^{(i}H^{jk)}=0
\label{6.10}
\eea
and has the super-Weyl transformation law
\bea
\d_\s H^{ij}=3\s H^{ij}
~.\eea
It is assumed that $H^{ij}$ is nowhere vanishing, 
 $H^2 := \hf H^{ij}H_{ij} >0$. 
The super-Weyl gauge freedom may be used  to impose the gauge
condition
\bea
H^2=1\quad \Longleftrightarrow \quad
H^i{}_kH^k{}_j=- \d^i_j
~,
\label{linear-1}
\eea
which completely fixes
the super-Weyl invariance. 
Now  the analyticity constraint \eqref{6.10} 
and  gauge condition  \eqref{linear-1}  tell us that 
$H^{ij}$ is  annihilated by all the spinor covariant derivatives,
\bea
\cD_\hal^i H^{jk}=0
~.
\eea
This is consistent under the integrability condition
$\big\{ \cD_{\hal}^i , \cD_{\hbe}^j \big\} H^{kl}=0$, 
which leads to the following set of constraints:
\bsubeq \label{linear-000}
\bea
S^{ij}&=&S\,H^{ij}
~,~~~~~~
\label{linear-13} 
\\
N_{\ha\hb}&=&-X_{\ha\hb}
~,
\label{linear-3} \\
\cD_{\ha}H^{ij}&=& C_{\ha }{}^{k (i}H^{j)}{}_k
~,
\label{linear-2}
\eea
\esubeq
for some scalar superfield $S$.

\subsection{Supersymmetric backgrounds} \label{subsection7.1}

It is of interest to study those supersymmetric backgrounds
which support the curved superspace geometry just described.
All information about such backgrounds can be extracted from the results 
derived in sections 4 and 5 provided 
we take into account the additional conditions
\eqref{linear-1} -- \eqref{linear-000}.
The Killing spinor equation  \eqref{Killing-spinor3/2} 
turns into 
\bea
\bD_\ha\e^k
&=&
\Big(\,
{1\over 2} \,s\,h^k{}_l\G_\ha
+{1\over 2} \d_l^k x_{\ha\hb}\G^\hb
+{1\over 8}\d_l^k\ve_{\ha\hb\hc\hd\he}x^{\hb\hc}\S^{\hd\he}
-{1\over 2}c^\hb{}^k{}_l\S_{\ha\hb}
\Big)
\e^l
~,~~~~~~
\label{7.22}
\eea
where we have introduced the component fields 
\bea
s:=S|~,\qquad
h^{ij}:=H^{ij}|
\eea
and used the component relations 
\bea
 s^{ij}=s\,h^{ij}~, \qquad n_{\ha\hb}=-x_{\ha\hb}~, \label{eq7.24}
 \eea
 which follow from  \eqref{linear-13} and \eqref{linear-3}.
The isovector field is constrained by 
\bea
h^i{}_kh^k{}_j=- \d^i_j ~, \qquad
 \bD_{\ha}h^{ij}= c_\ha{}^{k(i} h^{j)}{}_k~. \label{eq7.25}
\eea

\subsection{Supersymmetric backgrounds with eight supercharges}
\label{section7.2}

Different maximally supersymmetric backgrounds  appear 
depending on whether the  fields 
$s$ and/or  $c_\ha{}^{ij}$ are zero or not.
 In fact, there are three cases: 
(i) $s\ne 0$ ; (ii) $c_\ha{}^{ij} \ne s = 0$; and $s = c_\ha{}^{ij} =  0$. They 
correspond to those worked out in sections \ref{section5.1}, \ref{section5.2} and \ref{section5.3}, 
respectively. The choice of the real $\cO(2)$ multiplet compensator requires that we take into account 
the additional relations \eqref{eq7.24} and \eqref{eq7.25}. It is then 
straightforward to read off the curvatures 
and Weyl tensors from the corresponding ones in section \ref{8-susy}.

\section{Off-shell supergravity}
\setcounter{equation}{0}

We turn to an off-shell formulation for 5D minimal supergravity
obtained by coupling the Weyl multiplet to the following compensators:
(i) the vector multiplet; and (ii) the $\cO(2)$ multiplet.  
This is the 5D analogue of the off-shell formulation for 4D $\cN=2$
supergravity proposed by de Wit, Philippe and Van Proeyen \cite{deWPV}.
Our goal is to elucidate  those restrictions on the supersymmetric backgrounds
that follow from the structure of the compensators chosen. 

As has been discussed above, the super-Weyl gauge freedom may be fixed 
using one of the two compensators, either by imposing the condition 
$W=1$ or the alternative one  $H=1$.
To start with, we do not impose any super-Weyl condition and list 
those off-shell relations which turn into non-trivial constraints upon  
imposing a super-Weyl gauge.

In the case of the vector compensator, 
the Bianchi identity \eqref{W-BI} can be interpreted 
as an equation that expresses $C_\ha{}^{kl}$ in terms of $W$:
\begin{subequations}\label{8.1+2}
\bea
 C_{\ha}{}^{ij} 
 =
 \frac{\ri}{2W}(\G_\ha)^{\hal\hbe}\cD_{\hal}^{(i}\cD_{\hbe}^{j)}W
~.~~~~~~
\label{8.1}
\eea
It is also useful to rewrite equation \eqref{FSb} as
\bea
X_{\ha\hb}
&=&
\frac{1}{W}\Big(
F_{\ha\hb}
-{\ri\over 4}(\S_{\ha\hb})^{\hga\hde}\cD_{\hga}^k\cD_{\hde k}W\Big)~.
\label{8.2}
\eea
\end{subequations}
The relation expresses the torsion superfields $X_{\ha\hb}$ in terms of 
the vector multiplet.
In the super-Weyl gauge $W=1$, the relations 
\eqref{8.1+2}
take the form \eqref{minimal-2}.

In the case of the $\cO(2)$ compensator, 
the off-shell constraint on $H^{ij}$, eq. \eqref{6.10}, implies the following relations:
\begin{subequations} \label{8.3}
\bea
 X_{\ha\hb} + N_{\ha\hb} 
&=&
\frac{\ri}{4}(\S_{\ha\hb})^{\hal \hbe} H^{\hf} \cD_\hal^k \cD_{\hbe k} H^{-\hf} 
~;
%%%%%%
\\
S^{(i}{}_{k} H^{j)k}
&=& 
-\frac{\ri}{48 H^2}H^{(i}{}_{k}\Big(
 \cD^{\hal j)} \cD_{\hal}^{k} H^2 
-2 (\cD^{\hal j)}H)\cD_{\hal}^{k} H 
\Big)
~;
%%%%%%
\\
C_{\ha}^{(i}{}_{k} H^{j)k}
& =& 
-\cD_{\ha} H^{ij} 
- \frac{\ri}{16H^2} (\G_\ha)^{\hal\hbe} H^{(i}{}_{k}\Big(
\cD_\hal^{j)} \cD_{\hbe}^{k} H^2 
-2(\cD_{\hal}^{j)} H) \cD_{\hbe}^{k} H 
\Big)
~.
\eea
\end{subequations}
The first relation completely determines $X_{\ha\hb} + N_{\ha\hb}$
in terms of $H^{ij}$.
In the super-Weyl gauge $H=1$, the relations \eqref{8.3}
reduce to \eqref{linear-000}.

\subsection{Supersymmetric backgrounds}

Looking at the relations \eqref{8.1+2} and \eqref{8.3}, it appears that 
the super-Weyl gauge $H=1$ is simpler to deal with. This gauge condition 
and its implications, worked out in section  \ref{section7}, will be used in the 
remainder of this section. 
We have to analyse the implications of the supersymmetry invariance of $W$, 
\bea
\x^{\hat{A}}\cD_{\hat{A}}W
=0
~.
\label{inv-vector}
\eea
As before, we are interested in purely bosonic backgrounds,
and thus we require
\bea
\cD_\hal^i W| =  0
~.
\label{8.4}
\eea
Demanding this condition to be supersymmetric, $\d (\cD_\hal^i W)|= 0$, gives
\bea
\e^\hal_i\Big{[}
\ve_{\hal\hbe}\ve^{\hga\hde}\cD_{\hga}^{(i}\cD_{\hde}^{j)}
+(\G^\ha)_{\hal\hbe}(\G_\ha)^{\hga\hde} 
\Big( \cD_{\hga}^{(i}\cD_{\hde}^{j)} 
&-& \ri \ve^{ij} \cD_{\hal\hbe} \Big) \non \\
&+&\ve^{ij}(\S^{\ha\hb})_{\hal\hbe}(\S_{\ha\hb})^{\hga\hde}\cD_{\hga}^{k}\cD_{\hde k}
\Big{]} W| =0
~.
~~~~~~~
\eea
This  is equivalent to
\bea
\Big{[}
y^{ij}
\mathbbm{1}
+2
{\rm w} c_\ha{}^{ij} \,\G^\ha
-4\ve^{ij} \G^\ha {\bf D}_\ha {\rm w} 
+4\ve^{ij}\big(f_{\ha\hb}-{\rm w} x_{\ha\hb}\big)
\S^{\ha\hb}
\Big{]} \e_{ j} =0
~,
\label{eq-poinc-AdS-off-shell}
\eea
where we have introduced the component fields 
\bea
{\rm w}:=W| ~, \qquad y^{ij} := \ri \cD^{\hga(i}\cD_{\hga}^{j)}W|~.
\eea
By construction, the scalar $\rm w$ is nowhere vanishing. 
Eq. \eqref{eq-poinc-AdS-off-shell} is the additional condition
on any supersymmetric background, which 
comes from the vector compensator.  
The other conditions are given in subsection \ref{subsection7.1}.

%%%%%%%%%%%%%%%%%%%%%%%%%%%%%%%
%%%%%%%%%%%%%%%%%%%%%%%%%%%%%%%%

\subsection{Supersymmetric backgrounds with eight supercharges}

In the case of maximally supersymmetric backgrounds, 
equation
\eqref{eq-poinc-AdS-off-shell} is solved by
\bea
{\rm w} = {\rm const}~, ~~~y^{ij} =0
~,~~~
c_\ha{}^{ij}=0
~,~~~
x_{\ha\hb}=\frac{1}{\rm w}f_{\ha\hb}
~.
\label{8.8}
\eea
It should be kept in mind that  
the two-form
$f:=\hf f_{\ha\hb} e^\ha e^\hb $ is a U(1) field strength, and hence
it is closed, 
$\rd f = 0$. 

Since we  consider the maximally supersymmetric backgrounds, 
it follows from eq. \eqref{8.4} that 
\bea
\cD_\hal^i W=0\quad \Longrightarrow \quad W = {\rm const}~.
\eea
The first and second conditions in \eqref{8.8} are corollaries of this result.  From 
\eqref{8.1} we also deduce
\bea
C_\ha{}^{ij}=0~.
\eea

We can  now use the results of 
section \ref{8-susy}
to describe maximally supersymmetric backgrounds in off-shell supergravity. 
Note that in our case $c_\ha{}^{ij} =0$ and $n_{\ha\hb}=-x_{\ha\hb}$.

\subsubsection{The case $s\ne 0$}

When the scalar $s$ is nonzero,
all conclusions of subsection \ref{section7.2} hold.
In particular, the spacetime has  AdS$_5$ geometry.

\subsubsection{The case $s = 0$}

It remains to consider the case $s=0$.
Then
$s^{ij}=0 $ and $c_\ha{}^{ij}= 0$, 
and the geometry is formulated in terms of a single 
  two-form $x = \hf x_{\ha\hb} e^\ha e^\hb$ such that
\bea
\bD_\ha x_{\hb\hc}=
\frac{1}{8} \eta_{\ha[\hb}\ve_{\hc]\hd\he\hat{f}\hat{g}}x^{\hd\he}x^{\hat{f}\hat{g}}
~.
\label{8.11} 
\eea
This equation implies that the   two-form $x $ is closed, $\rd x=0$, 
which is consistent with the relation $x_{\ha\hb}=(1/{\rm w})f_{\ha\hb}$.

\section{Supersymmetric solutions in Poincar\'e and anti-de Sitter supergravities}
\setcounter{equation}{0}

In sections 6 and 7, we studied the restrictions on supersymmetric backgrounds
that originate due to the presence of a single conformal compensator. 
In section 8 we considered the off-shell supergravity formulation
obtained by coupling the Weyl multiplet to two
compensators: (i) the vector multiplet; and (ii) the $\cO(2)$ multiplet.  
It was demonstrated that the presence of a second compensator leads to 
additional restrictions on  supersymmetric backgrounds.
Now we turn to analysing supersymmetric solutions 
in  this supergravity theory,  with or without  a cosmological term.
 Our analysis will be restricted to the case of on-shell supergravity backgrounds.

It may be shown that the supergravity equations of motion\footnote{Similar 
equations of motion occur in 4D $\cN=2$ 
(gauged)  supergravity \cite{BK10,BK11}.}  
are
\bsubeq \label{eq-mot}
\bea
H-W^3&=&0~,
\label{eq-mot-0}
\\
{\mathbb H}^{ij}+\c H^{ij}&=&0~,
\label{eq-mot-1}
\\
{\mathbb W}+3\c W&=&0~,
\label{eq-mot-2}
\eea
\esubeq
with $\c$ the cosmological constant. 
Here ${\mathbb H}^{ij}$  is the composite $\cO(2)$ multiplet \eqref{mbbH-def},
and ${\mathbb W}$ is a composite vector multiplet constructed out of the $\cO(2)$ compensator. The latter is defined by 
\bea
\mathbb W = \bar{\mathbb W}
&= &
 \frac{\ri}{4} H \big(\cD^{ij}+12\ri S^{ij}\big) \Big(\frac{H_{ij}}{H^2}\Big)
\eea
and 
obeys the Bianchi identity
\bea
\cD_{\hal}^{(i}\cD_{\hbe}^{j)}\mathbb W
-{1\over 4}\ve_{\hal\hbe}\cD^{\hga(i}\cD_{\hga}^{j)} \mathbb W 
= {\ri\over 2} C_{\hal\hbe}{}^{ij} \mathbb W ~.~~~~~~
\eea
Its super-Weyl transformation law is 
\bea
\d_\s  \mathbb W=\s \mathbb W~.
\eea

Let us comment on the equations of motion \eqref{eq-mot}. 
The supergravity theory is described
in terms of three interacting multiplets: (i) the Weyl multiplet; (ii) the vector multiplet; and 
(iii) the $\cO(2) $ multiplet. It may be shown that, modulo gauge freedom,  the Weyl multiplet is described by a 
single unconstrained real prepotential $G$.\footnote{This 
can be done in complete analogy with the case of 4D $\cN=2$ supergravity \cite{KT}.}
The equation \eqref{eq-mot-0} is obtained by varying the supergravity action with respect 
to $G$. The meaning of \eqref{eq-mot-0} is that the supercurrent of pure supergravity 
is equal  to zero. 

In general, given a super-Weyl invariant theory of dynamical (matter) superfields $\vf^i$ coupled
to the Weyl multiplet, the supercurrent of this theory is a real scalar superfield defined by 
\bea
\cT = \frac{\D }{\D G}   S[\vf ] ~,
\eea
where $\D / {\D G}$ denotes a covariantized variational derivative with respect to $G$.
The supercurrent turns out to obey the conservation 
equation\footnote{The  supercurrent multiplet in 5D $\cN=1$ Poincar\'e supersymmetry
was introduced by Howe and Lindstr\"om \cite{HL}.} 
\bea
\big(\cD^{ij}+12\ri S^{ij}\big)\cT=0
\label{8.6}
\eea
provided the dynamical superfields obey their equations of motion, 
$\d S[\vf ] / \d \vf^i = 0$. 
The super-Weyl transformation law of $\cT$ is 
\bea
\d_\s \cT = 3\s \cT~,
\eea
which makes the equation \eqref{8.6} super-Weyl invariant. 
It is an instructive exercise to prove that the left-hand side of  \eqref{eq-mot-0}
obeys the constraint 
\bea
\big(\cD^{ij}+12\ri S^{ij}\big)(H-W^3) =0
\eea
provided the equations  \eqref{eq-mot-1} and  \eqref{eq-mot-2} hold. 

 The equations of motion  \eqref{eq-mot-1} and  \eqref{eq-mot-2} correspond 
to the vector and $\cO(2$) compensators, respectively. 
 The derivation of  these equations will be given elsewhere. 

Note that we can always choose the super-Weyl gauge \eqref{linear-1},
\bea
H=1~.
\eea
As shown in section \ref{section7}, 
this gauge condition implies   
\bea
\cD_\hal^i H^{jk} = 0~, \qquad 
S^{ij}=SH^{ij}~,
\eea 
for some scalar superfield $S$. 
Due to the equation of motion \eqref{eq-mot-0}, 
the field strength  $W$ also becomes constant, 
\bea
W=1 ~.
\eea
Moreover, both eqs. \eqref{eq-mot-1} and \eqref{eq-mot-2} become equivalent to
\bea
S
&=&
\frac{1}{2} \c~.
\eea

Since $W=1$ and $H=1$ on the mass shell, it holds that
\bea
C_\ha{}^{kl}=
0~,\qquad
X_{\ha\hb}=F_{\ha\hb}=-N_{\ha\hb}
~.
\label{9.13}
\eea
Due to \eqref{linear-2}, $H^{kl}$ is actually covariantly constant, 
\bea
\cD_{\hat{A}}H^{kl}=0~,
\eea
and therefore the SU(2) curvature factorizes, 
\bea
R_{\hat{A}\hat{B}}{}^{kl} =
 R_{\hat{A}\hat{B}} H^{kl}~,
\eea
for a  closed super two-form $  R_{\hat{A}\hat{B}} $ given by
\bea
R_{\hal}^i{}_{\hbe}^j
=
\frac{3\ri}{2}\chi\ve_{\hal\hbe}\ve^{ij}
~,~~~~~~
%%%%%%
R_\ha{}_{\hbe}^j
=
0~,~~~~~~
%%%%%%
R_{\ha\hb}
=
- \frac{3}{4}\chi F_{\ha\hb}
~.
\label{8.15}
\eea
This  super two-form proves to be proportional to the U(1) field strength 
$F_{\hat{A}\hat{B}} $, eq. \eqref{FS6.3},
\bea
R_{\hat{A}\hat{B}} = - \frac{3 }{4} \c F_{\hat{A}\hat{B}} ~.
\label{8.17}
\eea
Now  the local SU(2) symmetry may be used to choose
the corresponding connection in the form 
$\F_{\hat{A}}{}^{kl}=  \F_{\hat{A}}H^{kl}$.
As a result, the SU(2) group
reduces to a U(1) subgroup generated by $\cJ:=-\ri H^{kl}J_{kl}$.
Due to \eqref{8.17}, we may identify (up to a factor)  $  \F_{\hat{A}}$ 
with the U(1) connection $  \cV_{\hat{A}}$ in \eqref{6.1}.

For on-shell supergravity under consideration, 
we are interested in backgrounds 
that possess some rigid supersymmetry.  
Using the gauge conditions described, 
the Killing spinor equation \eqref{6.8} turns into
\bea
{\bD}_\ha\e^k
&=& \frac{1}{4}  \c \G_\ha h^k{}_l\e^l 
+ \frac{1}{ 8}\d_l^kf_{\hb\hc}\Big(
\ve_{\ha}{}^{\hb\hc\hd\he}\S_{\hd\he}
+4\d^\hb_\ha\G^\hc
\Big)
\e^l
~.
\label{Killing_on-shell}
\eea

The Killing spinor equation \eqref{Killing_on-shell} coincides with the 
one derived in \cite{GG2}. In the case of Poincar\'e supergravity, $\c=0$,
it reduces to the Killing spinor equation given in \cite{GG1}.
The supersymmetric backgrounds for on-shell 
simple Poincar\'e and anti-de Sitter supergravity theories 
in five dimensions have been studied
in detail in \cite{GG1} and \cite{GG2}, respectively.
There is no  need to repeat here the analysis given there. 

In the case of anti-de Sitter supergravity, 
$\c \neq 0$, the isovector $s^{ij}$ is non-zero, $s^{ij} = \hf  \c h^{ij}$.  
Then our earlier analysis implies that  AdS${}_5$ is the only maximally supersymmetric solution. 
This agrees with the conclusions of \cite{GG2}.

\section{Concluding comments}
\setcounter{equation}{0}

In this paper we have developed the formalism 
to construct off-shell supersymmetric backgrounds
within the superspace formulation for 5D conformal supergravity 
\cite{KT-Mconfsugra5D}. 
For those superspace backgrounds which obey the equations of motion for 
Poincar\'e or anti-de Sitter supergravity, we have naturally reproduced 
the supersymmetric solutions constructed in \cite{GG1,GG2}.

Although we presented a number of supersymmetric backgrounds, 
a classification of such semi-Riemannian spaces was not our goal. 
Given a semi-Riemannian space that admits at least one rigid supersymmetry, 
our ultimate aim was to embed it in a curved background superspace such that 
its geometry is of the type described in section 2.
After that it becomes trivial to generate rigid supersymmetric theories 
on this space by making use of the off-shell supergravity-matter systems
presented in \cite{KT-Msugra5D,KT-Msugra5D2,KT-Mconfsugra5D}.
In this sense, the curved superspace approach 
is much more powerful than 
the Noether procedure advocated, e.g., in \cite{FS}.

To illustrate the power of the curved superspace approach at generating rigid
supersymmetric theories, it suffices to consider the example of 5D anti-de Sitter space.
Eight years ago, two of us \cite{KT-M07}
constructed the most general off-shell supersymmetric nonlinear $\s$-models in 
5D $\cN=1$ AdS superspace
formulated in terms of covariant weight-zero polar hypermultiplets. 
A year later, the construction of  \cite{KT-M07}
was extended to the case of 4D $\cN=2$ AdS supersymmetry \cite{KT-M08}.
However, since the $\s$-models proposed in  \cite{KT-M07,KT-M08} made use of off-shell 
supermultiplets with infinitely many auxiliary fields, which have never been dealt with in the framework 
of superconformal tensor calculus, these theories remained largely unnoticed. 
In 2011, two separate developments took place. The most general nonlinear $\s$-models 
with 4D $\cN=2$ AdS and 5D $\cN=1$ AdS 
supersymmetries were constructed in   \cite{BKads} and 
\cite{BaggerXiong,BaggerLi}, 
respectively, in terms of 4D $\cN=1$ chiral superfields.\footnote{The component formulation 
of the 5D $\cN=1$ supersymmetric $\s$-models constructed in  \cite{BaggerXiong} was given in \cite{BaggerLi}.}
The common feature of the 4D $\cN=2$ and 5D $\cN=1$ AdS supersymmetries is 
that the $\s$-model target spaces
are those hyperk\"ahler manifolds which possess a Killing vector field 
generating an SO(2) group of rotations on the two-sphere of complex structures.\footnote{Such
hyperk\"ahler manifolds were first described in \cite{HitchinKLR}.}
Not all hyperk\"ahler manifolds possess such an SO(2) isometry group. 
This clearly differs from the 4D $\cN=2$ or 5D $\cN=1$ Poincar\'e supersymmetries
where  arbitrary hyperk\"ahler manifolds can originate
as target spaces of  supersymmetric $\s$-models \cite{A-GF,HKLR}.
In 2012, Ref. \cite{BKLT-M}  established the one-to-one correspondence 
between the two types of $\cN=2$ supersymmetric $\s$-models in AdS$_4$:  the off-shell
 \cite{KT-M08} and the on-shell \cite{BKads} ones.   
Similar considerations may be used to establish a one-to-one correspondence 
between the $\cN=1$ supersymmetric $\s$-models in AdS$_5$ constructed
in \cite{KT-M07} and  \cite{BaggerXiong}.

The off-shell supersymmetric $\s$-models with eight supercharges 
in AdS$_4$ \cite{KT-M08} and AdS$_5$ \cite{KT-M07} 
are constant-curvature deformations 
of the family of $\cN=2$ rigid supersymmetric $\s$-models 
in ${\mathbb R}^{3,1}$ introduced 
in \cite{K98} and studied in \cite{GK1,GK2} (see also \cite{KL} 
for the 5D $\cN=1$ extension).\footnote{The supersymmetric $\s$-models 
introduced in \cite{K98} form a special subfamily in the general family of 
polar multiplet $\s$-models pioneered by Lindstr\"om and Ro\v{c}ek \cite{LR}.}
The target space $\cM$ of such a nonlinear  $\s$-model 
was shown  in \cite{K98,GK1,GK2}
to be an open domain of the zero section 
of the cotangent bundle $T^*\cX$  of a real analytic K\"ahler manifold $\cX$
(the off-shell $\s$-model action \cite{K98} 
is constructed in terms of the K\"ahler potential $K(\F , \bar \F )$ of $\cX$).
Since the target spaces of any 4D $\cN=2$ rigid supersymmetric $\s$-models 
are hyperk\"ahler \cite{A-GF}, $\cM$ is a hyperk\"ahler manifold, 
for any real analytic K\"ahler manifold $\cX$. 
Thus the superspace construction of  \cite{K98,GK1,GK2} provided a proof
that there exists a hyperk\"ahler structure on an open domain of  the zero section 
of the cotangent bundle $T^*\cX$  of a real analytic K\"ahler manifold $\cX$.
 This proof is much simpler than the ones given 
 in the mathematical literature \cite{Kaledin,Feix} 
 and appeared two years earlier than \cite{Feix}.\footnote{One of the authors of \cite{GK1}
(SMK)  was informed about Kaledin's work \cite{Kaledin} only after his talk, 
which was  given
 at the 32nd International Symposium Ahrenshoop on the Theory of Elementary Particles 	
(1-5 September 1998,  Buckow, Germany) and in which the results of \cite{GK1} were announced. 
Ref. \cite{GK2} is a written version of the talk given.}  
 For any real analytic K\"ahler manifold $\cX$, the off-shell 
 $\s$-model action of  \cite{K98,GK1,GK2}
possesses a U(1) rigid symmetry, which manifests in a certain U(1) isometry 
of the hyperk\"ahler space  $T^*\cX$. This U(1) isometry acts by scalar multiplication 
in the fibres and rotates the complex structures. This U(1) isometry group
of $T^*\cX$ plays an important role in  \cite{Kaledin,Feix}.

In the case of maximally supersymmetric backgrounds 
considered in subsection \ref{section5.2}, 
it is not difficult to construct a family of rigid supersymmetric 
$\s$-models as a generalization of the 
locally supersymmetric off-shell nonlinear $\s$-models given in 
 \cite{KT-Msugra5D,KT-Msugra5D2,KT-Mconfsugra5D}.
The dynamical variables of such a theory are 
a set of interacting {\it covariantly arctic weight-zero} multiplets 
$\U^I  $ and their smile-conjugates $ \breve{ \U}^{\bar I} $, 
and the dynamics is described  
by a  projective-superspace Lagrangian of the form 
\bea
\cL^{++} = C^{++}
{ K}({ \U}, \breve{ \U})~, \qquad C^{++} = C^{ij}u^+_i u^+_j ~,
\eea
where ${K}(\F^I, {\bar \F}^{\bar J}) $ is the K\"ahler potential of 
a real analytic K\"ahler manifold $\cM$, and $u^+_i$ are 
homogeneous complex coordinates for ${\mathbb C}P^1$.  
The supersymmetric action constructed from $\cL^{++} $
proves to be  invariant under K\"ahler transformations of the form
\be
{K}({ \U}, \breve{ \U})~\to ~{ K}({ \U}, \breve{ \U})
+{ \L}({ \U}) +{\bar { \L}} (\breve{ \U} )~,
\ee
with ${ \L}(\F^I)$ a holomorphic function.
 It is of interest to understand the target-space geometry of such nonlinear 
 $\s$-models, in particular its dependence on the 5-vector parameter 
 $C^\ha$ of the curved superspace under consideration. 
 
It appears that only superconformal $\s$-sigma models can be consistently  
 defined in the case of those maximally supersymmetric backgrounds 
in  subsection  \ref{section5.3} that are characterized by the condition 
$X_{\ha\hb}+N_{\ha\hb}\ne0$,
 because the holonomy group
of the superspace \eqref{5.46} then includes  the $R$-symmetry group SU(2). 
However, if  $X_{\ha\hb}+N_{\ha\hb}=0$, the SU(2) curvature is identically zero.  
 
In our discussion of 5D supersymmetric backgrounds, the bosonic conditions
\eqref{Ddim1} were postulated. Actually such conditions naturally originate as 
consistency requirements for the existence of rigid supersymmetry
transformations.  Indeed, let $\cT $ be any  bosonic component of the superspace torsion and curvature 
tensors in \eqref{algebra}, which correspond to a supersymmetric background.  
The variation of $\cT$ under a rigid supersymmetry 
transformation must vanish, and hence
\bea
0 = \e^\hal _i \,\cD^i_\hal \cT |~,
\eea
 where we have made use of the conditions 
 $K^{\ha \hb}| =0$, $K^{ij}|=0$ and $\s [\x]=0$. For this to hold, it suffices
 to require the spinor component $ \cD^i_\hal \cT |$ to vanish,
 $\cD^i_\hal \cT |=0$. 
 On the other hand, if we are only interested in those backgrounds that 
possess conformal supersymmetries, it is not necessary to impose 
the bosonic conditions \eqref{Ddim1}. To see this, let us start from 
a purely bosonic background
possessing a conformal supersymmetry 
and then introduce a conformally related superspace 
defined by \eqref{CRSG}.  For the latter superspace, 
the requirements \eqref{4.100} still hold, but some of the conditions 
$K^{\ha \hb}| =0$, $K^{ij}|=0$ and $\s|=0$  are no longer true. 
Moreover, some fermionic components of the superspace torsion 
and curvature tensors may be non-zero.   
 
Recently, there have appeared two publications devoted to supersymmetric 
backgrounds for 5D $\cN=1$ supergravity with Euclidean signature \cite{Pan,IM}. 
Our conformal Killing equation \eqref{Killing-spinor3/2} is analogous to those
given in \cite{Pan,IM}. 
\\

\noindent
{\bf Acknowledgements:}\\
We are grateful to Daniel Butter for valuable suggestions and comments
on the manuscript. 
This work is supported in part by the Australian Research Council
projects DP1096372, DE120101498 and DP140103925.

\appendix

\section{(Conformal) isometries in curved space}
\setcounter{equation}{0}

In this appendix we recall how the problem of computing the (conformal) isometries 
of a curved spacetime is addressed within the Weyl-invariant formulation 
for gravity \cite{Deser}. 
Our presentation follows \cite{K-Mool}.

We start by recalling three known approaches to  the description of gravity 
in $d$ dimensions: (i) metric formulation; 
(ii) vielbein formulation; and (iii) Weyl-invariant formulation. 
In the standard metric approach, the gauge field is a metric tensor $g_{mn} (x) =g_{nm}(x) $
constrained to be nonsingular,  $g:= \det (g_{mn}) \neq 0$. The gauge transformation is 
\bea
 \d g_{mn} =  \nabla_m \x_n + \nabla_n \x_m~,
 \eea
 with  the gauge parameter $\x = \x^m (x) \pa_m$ 
 being a vector field generating an infinitesimal diffeomorphism.
 
In the vielbein formulation,
the gauge field is  a vielbein $e_m{}^a (x)$ that constitutes a basis in the tangent space at $x$, 
for any spacetime point $x$, 
  $e:=\det (e_m{}^a ) \neq 0$.
The metric becomes a composite field defined by $g_{mn} = e_m{}^a e_n{}^b \eta_{ab} $.
The gauge group is now larger than in the metric approach. 
It includes general coordinate and local Lorentz transformations, 
\bea
\d \nabla_a = [ \x^b \nabla_b +\hf K^{bc}M_{bc} , \nabla_a]~,
\eea
with the gauge parameters $\x^a (x) = \x^m (x) e_m{}^a (x) $ and $K^{ab} (x) = - K^{ba}(x)$ 
being completely arbitrary.
The gauge transformation makes use of the   torsion-free
covariant derivatives 
\bea
\nabla_a = e_a{}^m  \pa_m +\hf \o_a{}^{bc}  M_{bc} ~, \qquad 
[\nabla_a , \nabla_b ] = \hf R_{ab}{}^{cd} M_{cd} ~.
\eea
Here $M_{bc} = -M_{cb}$ denotes the Lorentz generators, 
$e_a{}^m $ the inverse vielbein, $e_a{}^m e_m{}^b = \d_a{}^b$,  
and $\o_a{}^{bc}$ the torsion-free Lorentz connection. 

As is well known, the torsion-free constraint
\bea
T_{ab}{}^c=0 \quad \Longleftrightarrow \quad 
[\nabla_a , \nabla_b ] \equiv T_{ab}{}^c \nabla_c +\hf R_{ab}{}^{cd} M_{cd} 
=\hf R_{ab}{}^{cd} M_{cd}
\eea
is invariant under  Weyl (local scale) transformations 
\bea
\nabla_a \to \nabla'_a =  \re^{\s} \Big( \nabla_a +(\nabla^b\s) M_{ba}\Big)~,
\label{A.5}
\eea
with the parameter $\s(x)$ being completely arbitrary.  
This transformation is induced by that of the gravitational field 
\bea
 e_a{}^m \to \re^\s e_a{}^m 
\quad \Longrightarrow \quad g_{mn} \to \re^{-2\s}g_{mn} ~.
\eea
Most field theories in curved space do not possess Weyl invariance.
In particular, the  pure gravity action with  a cosmological term
\bea
S= \frac{1}{2\k^2} \int \rd^d x \,e \,  R   -\frac{\L}{\k^2} \int \rd^dx \,e 
\label{A.9}
\eea
is not invariant under the Weyl transformations \eqref{A.5}.

Weyl-invariant matter theories are curved-space extensions of ordinary conformally invariant theories. 
As an example, consider the model for a scalar field $\vf$ with action
\bea
S=-\hf \int \rd^d x \,e \,\Big\{ \nabla^a \vf \nabla_a \vf +\frac{1}{4} \frac{d-2}{d-1} 
R \vf^2  +{\l} \vf^{2d/(d-2)} 
\Big\}~,
\label{A.7}
\eea
with $R$ the scalar curvature and $\l$ a coupling constant. The action is Weyl invariant\footnote{The
Weyl transformation law of $R$ is $ R\to \re^{2\s} \Big\{ R + 2(d-1) \nabla^a\nabla_a \s 
- (d-2)(d-1) ( \nabla^a \s ) \nabla_a \s\Big\}$.}
provided 
$\vf$ transforms as 
\bea 
\vf \to \vf' = \re^{\hf (d-2)\s} \vf~. 
\label{A.8} 
\eea

In the Weyl-invariant formulation for gravity,
the gravitational field is described in terms of two gauge fields. 
One of them is the vielbein $e_m{}^a  (x)$ and the other is a conformal compensator $\vf (x)$
with the Weyl transformation law \eqref{A.8}. 
Unlike the matter model \eqref{A.7}, the compensator is constrained to be nowhere 
vanishing, $\vf \neq 0$.  The gravity gauge group is defined to include the general coordinate,
local Lorentz and Weyl transformations\begin{subequations}\label{A.10}
\bea
\d \nabla_a &=&  [ \x^b \nabla_b +\hf K^{bc}M_{bc} , \nabla_a]
+  \s \nabla_a +(\nabla^b\s) M_{ba} \equiv ( \d_\cK + \d_\s )\nabla_a~,  \label{A.10a} \\
\d \vf &=&  \x^b \nabla_b \vf + \hf (d-2)\s \vf \equiv ( \d_\cK + \d_\s ) \vf~ , 
\eea
\end{subequations}
where we have denoted $\cK:= \x^b \nabla_b +\hf K^{bc}M_{bc} $.
In this approach, any dynamical system is required to be invariant under
 the general coordinate, local Lorentz and Weyl transformations. 
 In particular, the Weyl-invariant gravity action is 
 \bea
S=\hf \int \rd^d x \,e \,\Big\{ \nabla^a \vf \nabla_a \vf +\frac{1}{4} \frac{d-2}{d-1} 
R \vf^2  +{\l} \vf^{2d/(d-2)} 
\Big\}~.
\label{A.11}
\eea
Applying a finite Weyl transformation allows us to choose a gauge 
\bea 
\vf = \frac{1}{2\k}  \sqrt{\frac{d-1}{d-2}}~,
\eea
in which the action turns into \eqref{A.9}.

A vector field $\x = \x^m \pa_m = \x^a e_a$, with $e_a := e_a{}^m \pa_m$, 
is  conformal Killing if there exist local Lorentz $K^{bc}[\x]$ and 
Weyl $\s [\x] $ parameters such that 
\bea
 \Big[ \x^b \nabla_b +\hf K^{bc} [\x] M_{bc} , \nabla_a\Big]
+  \s [\x] \nabla_a +(\nabla^b\s [\x]) M_{ba} =0~.
\eea
A short calculation gives
\bea
K^{bc} [\x] &=& \hf \big( \nabla^b \x^c - \nabla^c \x^b \big)~, \qquad
\s[\x] = \frac{1}{d} \nabla_b \x^b  
\eea
as well as the conformal Killing equation
\bea
 \nabla^a \x^b + \nabla^b \x^a = 2 \eta^{ab} \s[\x] ~.
 \eea
The set of all conformal Killing vector fields of a given spacetime
is a finite-dimensional Lie algebra with respect to the standard Lie bracket
for vector fields. It is the conformal algebra of the spacetime.

Two spacetimes $(\nabla_a, \vf)$ and 
$(\widetilde{\nabla}_a,  \widetilde{\vf})$ are said to be conformal 
if their covariant derivatives are related to each other as follows:
\bea
\widetilde{\nabla}_a &=&  \re^{\r} \Big( \nabla_a +(\nabla^b\r) M_{ba}\Big) ~,\qquad
\widetilde{\vf} = \re^{\hf (d-2)\r} \vf ~,
\label{A.16}
\eea
for some $\r$. These spacetimes 
have the same conformal Killing vector fields 
$\x = \x^a e_a = \tilde{\x}^a \tilde{e}_a$. 
The parameters $K^{cd}  [\tilde{\x} ] $ and $\s [ \tilde{\x} ] $ are related to 
$K^{cd}  [{\x} ] $ and $\s [ {\x} ] $ as follows:
\bea
\cK [\tilde{\x} ] &:=& \tilde{\x}^b \widetilde{\nabla}_b + \hf K^{cd}[\tilde{\x} ] M_{cd}
=\cK[\x]~, \\
\s [ \tilde{\x} ] &=& \s [ \x ] - \x \r ~.
\eea

A vector field $\x = \x^m \pa_m = \x^a e_a$, with $e_a := e_a{}^m \pa_m$, 
is   Killing if there exist local Lorentz $K^{bc} [\x]$ and 
Weyl $\s [\x] $ parameters such that 
\begin{subequations}\label{A.19}
\bea
 \Big[ \x^b \nabla_b +\hf K^{bc} [\x] M_{bc} , \nabla_a\Big]
+  \s [\x] \nabla_a +(\nabla^b\s [\x]) M_{ba} &=&0 ~, \\
\x \vf +\hf (d-2) \s[\x] \vf &=&0 ~.
\eea
\end{subequations}
The set of all conformal Killing vector fields of a given spacetime
is a finite-dimensional Lie algebra. By construction, it is a subalgebra of
the conformal algebra of the spacetime.   
The Killing equations \eqref{A.19} are Weyl invariant in the following sense.
Given a conformally related spacetime $(\widetilde{\nabla}_a,  \widetilde{\vf})$
defined by eq.~\eqref{A.16},
the  Killing equations \eqref{A.19} have the same functional form when rewritten in 
terms of  $(\widetilde{\nabla}_a,  \widetilde{\vf})$. In particular, 
\bea
\x \widetilde{\vf} + \hf (d-2)\s[\tilde{\x }] \widetilde{\vf} =0~.
\eea

Due to Weyl invariance, we can work with a conformally related spacetime such that 
\bea
 \vf =1~.
 \eea
Then for $d >2$ the Killing equations turn into 
\bea
 \Big[ \x^b \nabla_b +\hf K^{bc} [\x] M_{bc} , \nabla_a\Big]  &=&0~, \qquad
 \s[\x] = 0 ~.
 \eea
This is equivalent to the standard Killing equation
\bea
 \nabla^a \x^b + \nabla^b \x^a = 0~.
 \eea
 
 %%%%%%%%%%%%%%%%%%%%%%%%%%%%%%%%%%%%%%%%%%%%%%%%%%%
 %%%%%%%%%%%%%%%%%%%%%%%%%%%%%%%%%%%%%%%%%%%%%%%%%%%

\section{Conformal Killing spinors and bilinears}
\setcounter{equation}{0}

The famous classification of supersymmetric solutions \cite{GG1,GG2} in 5D $\cN=1$ 
Poincar\'e and anti-de Sitter supergravity theories was based on the use of 
the algebraic and differential properties of bilinears constructed from a Killing 
spinor. In this appendix we study the properties of such bilinears associated
with (conformal) Killing spinors in off-shell supergravity. 

Given a commuting spinor $\e_\hal^i$, we may construct the following real bilinears:
\bsubeq
\begin{align}
F &:= \e^k_\hga \e^\hga_k \ , \\
V_\ha &:= (\G_\ha)^{\hal\hbe} \e_\hal^k \e_{\hbe k} \ , \\
G_{\ha\hb}{}^{ij} &:= - (\S_{\ha\hb})^{\hal\hbe} \e_{\hal}^{(i} \e_{\hbe}^{j)} 
= G_{[\ha\hb ]}{}^{(ij)}
\ .
\end{align}
\esubeq
It is straightforward to show that the above bilinears satisfy the algebraic identities
\bsubeq \label{B.2}
\begin{align}
V^\ha V_\ha &= - F^2 \ , \label{B2.a}\\
\ve^{\ha\hb\hc\hd\he} G_{\hb\hc}{}^{ij} G_{\hd\he}{}_{kl} &= - \d^i_{(k} \d^j_{l)} V^\ha F \ , \\
V^\ha G_{\ha\hb}{}^{ij} &= 0 \ , \\
\ve^{\ha\hb\hc\hd\he} V_\hc G_{\hd\he}{}^{ij} &= 2 G^{\ha\hb}{}^{ij} F \ , \\
G_{\ha}{}^{\hc}{}^{ij} G_{\hc\hb}{}^{kl} &= \frac{1}{8} \ve^{k(i} \ve^{j)l} (\eta_{\ha\hb} F^2 + V_\ha V_\hb )
+ \frac{1}{4} \ve^{k(i} F G_{\ha\hb}{}^{j) l} + \frac{1}{4} \ve^{l (i} F G_{\ha\hb}{}^{j) k} \ , 
\label{B.2e}\\
V_{\hal\hbe} \e^{\hbe j} &= F \e_\hal^j \ , \\
G_{\hal\hbe}{}^{ij} \e^{\hbe k} &= - \hf \ve^{k (i} F \e_\hal^{j)} \ ,
\end{align}
\esubeq
where
\be 
V_{\hal\hbe} = (\G^\ha)_{\hal\hbe} V_\ha \ , \quad G_{\hal\hbe}{}^{ij} 
= \hf (\S^{\ha\hb})_{\hal\hbe} G_{\ha\hb}{}^{ij} \ .
\ee
Eq.~\eqref{B2.a} tells us that the five-vector $V^\ha$ is time-like or null.

Let $\e_\hal^i$ be a  conformal Killing 
spinor obeying the equation \eqref{conf-Killing-spinor}.
We then find the differential identities
\bsubeq
\begin{align}
\bD_\ha F &= \ri (\G_\ha)_\hal{}^\hbe \eta_\hbe^k \e^\hal_k + x_{\ha\hb} V^\hb + G_{\ha\hb}{}^{kl} c^{\hb}{}_{kl} \ , \\
\bD_\ha V_\hb &= \ri \eta_{\ha\hb} \eta^{\hga k} \e_{\hga k} - 2 \ri (\S_{\ha\hb})^{\hal\hbe} \eta_\hal^k \e_{\hbe k}
- s_{kl} G_{\ha\hb}{}^{kl} - \hf x_{\ha\hb} F \non\\
&\qquad- \frac{1}{8} \ve_{\ha\hb\hc\hd\he} n^{\hc\hd} V^\he 
+ \frac{1}{4} \ve_{\ha\hb\hc\hd\he} c^\hc{}^{kl} G^{\hd\he}{}_{kl} \ , \\
\bD_\ha G_{\hb\hc}{}^{ij} &= \frac{\ri}{2} \ve_{\ha\hb\hc\hd\he} (\S^{\hd\he})^{\hal\hbe} \eta_\hal^{(i} \e_\hbe^{j)} 
- \ri \eta_{\ha [\hb} (\G_{\hc]})^{\hal\hbe} \eta_{\hal}^{(i} \e_\hbe^{j)} \non\\
&\qquad - \hf \ve_{\ha\hb\hc\hd\he} s^{(i}{}_k G^{\hd\he j) k} - \hf \eta_{\ha [\hb} s^{ij} V_{\hc]} \non\\
&\qquad - \hf \ve_{\hd\he\hat{f} \hb\hc} x_\ha{}^\hd G^{\he\hat{f}}{}^{ij} 
+ \frac{1}{2} \ve_{\hd\he\hat{f}\ha [\hb} n^{\hd\he} G_{\hc]}{}^{\hat{f}}{}^{ij} \non\\
&\qquad- c_{[\hb}{}^{(i}{}_k G_{\hc] \ha}{}^{j) k} + \eta_{\ha [\hb} c^\hd{}^{(i}{}_k G_{\hc] \hd}{}^{j)k} \non\\
&\qquad + \hf \eta_{\ha [\hb} c_{\hc]}{}^{ij} F + \frac{1}{8} \ve_{\ha\hb \hc \hd\he} c^{\hd ij} V^\he \ .
\end{align}
\esubeq
These imply
\bsubeq
\bea 
&\hat{\bD}_{(\ha} V_{\hb)} = \frac{1}{5} \eta_{\ha\hb} \hat{\bD}^\hc V_\hc~, \label{B.5a}\\
& \hat{\bD}_{(\ha} G_{\hb) \hc}{}^{ij} = - \frac{1}{4} \eta_{\ha\hb} \hat{\bD}^\hd G_{\hc\hd}{}^{ij} 
+ \frac{1}{4} \eta_{\hc (\ha} \hat{\bD}^\hd G_{\hb) \hd}{}^{ij} \ , 
\label{B.4c}
\eea
\esubeq
where $ \hat{\bD}_\ha$ denotes the covariant derivative
\eqref{hatcovder}.
Eq. \eqref{B.5a} is equivalent to the conformal Killing equation \eqref{4.18}.

Now let us restrict  $\e_\hal^i$ to be  a Killing spinor, and hence
$\eta^\hal_k = 0$. Then we have 
\bsubeq
\begin{align}
\bD_\ha F &= x_{\ha\hb} V^\hb + G_{\ha\hb}{}^{kl} c^{\hb}{}_{kl}  \quad
\implies \quad V^\ha \bD_\ha F = 0 \ , \\
\bD_\ha V_\hb &= - s^{kl} G_{\ha\hb}{}_{kl} - \hf x_{\ha\hb} F - \frac{1}{8} \ve_{\ha\hb\hc\hd\he} n^{\hc\hd} V^\he 
+ \frac{1}{4} \ve_{\ha\hb\hc\hd\he} c^{\hc kl} G^{\hd\he}{}_{kl} 
\ .
\end{align}
and therefore $V^\ha$ is a Killing vector field, eq. \eqref{4.24}. 
Relation \eqref{B.4c} turns into
\begin{align}
\bD_\ha G_{\hb\hc}{}^{ij} &= 
- \hf \ve_{\ha\hb\hc\hd\he} s^{(i}{}_k G^{\hd\he j) k} - \hf \eta_{\ha [\hb} s^{ij} V_{\hc]} \non\\
&\qquad - \hf \ve_{\hd\he\hat{f} \hb\hc} x_\ha{}^\hd G^{\he\hat{f}}{}^{ij} 
+ \frac{1}{2} \ve_{\hd\he\hat{f}\ha [\hb} n^{\hd\he} G_{\hc]}{}^{\hat{f}}{}^{ij} \non\\
&\qquad- c_{[\hb}{}^{(i}{}_k G_{\hc] \ha}{}^{j) k} + \eta_{\ha [\hb} c^\hd{}^{(i}{}_k G_{\hc] \hd}{}^{j)k} \non\\
&\qquad + \hf \eta_{\ha [\hb} c_{\hc]}{}^{ij} F + \frac{1}{8} \ve_{\ha\hb \hc \hd\he} c^{\hd ij} V^\he \ .
\label{B.6c}
\end{align}
\esubeq
The last result implies
\begin{align} \bD_{[\ha} G_{\hb\hc]}{}^{ij} &= - \hf \ve_{\ha\hb\hc\hd\he} s^{(i}{}_k G^{\hd\he j) k} 
+ \frac{1}{8} \ve_{\ha\hb\hc\hd\he} c^\hd{}^{ij} V^\he - c_{[\ha}{}^{(i}{}_k G_{\hb \hc]}{}^{j) k} \non\\
&\qquad + \hf \ve_{\hd\he\hat{f} [\ha\hb} (n^{\hd\he} + x^{\hd\he}) G_{\hc]}{}^{\hat f ij} 
\label{8.7}
\end{align}
and
\be \hat{\bD}^\ha G_{\ha\hb}{}^{ij} = - s^{ij} V_\hb + c^\hd{}^{(i}{}_k G_{\hc\hd}{}^{j) k} + c_\hc{}^{ij} F \ .
\ee

Relation \eqref{8.7} dramatically simplifies if we are dealing with 
a supersymmetric solution of supergravity. 
In accordance with \eqref{9.13}, we then have
\bea
\bD_{[\ha} G_{\hb\hc]}{}^{ij} = - \hf \ve_{\ha\hb\hc\hd\he} s^{(i}{}_k G^{\hd\he j) k} ~,
\label{B.9}
\eea
where $s^{ij} = \hf \c h^{ij}$ is  covariantly constant. In the case of Poincar\'e 
supergravity, $\c =0$ and the right-hand side of \eqref{B.9} vanishes. 
Thus the three two-forms $G^{ij} := \hf G_{\ha\hb}{}^{ij} e^\ha e^\hb$ 
are closed,\footnote{Since the SU(2) curvature vanishes on-shell in Poincar\'e 
supergravity, the SU(2) connection can be completely gauged away.}
\bea
\rd G^{ij} =0~.
\eea
If the Killing vector $V^\ha$ is time-like, the closed two-forms $G^{ij}$ turn out 
to define  a hyper-K\"ahler structure on a 4D submanifold orthogonal to the orbit of 
$V^\ha$ \cite{GG1}. 

In the case of anti-de Sitter supergravity, $\c \neq 0$, we may introduce 
a two-form $G := s_{ij} \,G^{ij} $. In accordance with \eqref{B.9}, it is closed,
\bea
\rd G =0~.
\eea
${}$From eq.~\eqref{B.2e} we also deduce
\bea
G_{\ha}{}^{\hc}{} G_{\hc}{}^{\hb} 
= -\frac{1}{4} \c^2  (\d_\ha{}^\hb F^2 + V_\ha V^\hb )
\eea
If the Killing vector $V^\ha$ is time-like, the closed two-form $G$ proves 
to define  a K\"ahler structure on a 4D submanifold orthogonal to the orbit of 
$V^\ha$ \cite{GG2}.

%%%%%%%%%%%%%%%%%%%%%%%%%%%%%%%%
%%%%%%%%%%%%%%%%%%%%%%%%%%%%%%%%%%
%%%%%%%%%%%%%%%%%%%%%%%%%%%%%%%%%%%%%%%%%%%%%%%%%%%

\begin{footnotesize}

\end{footnotesize}

\end{document}